\documentclass[journal=jpcbfk,manuscript=article]{achemso}

\usepackage[version=3]{mhchem} % Formula subscripts using \ce{}
\usepackage{amssymb}
\usepackage{diagbox, eqparbox, hhline}
\usepackage{xcolor}
\usepackage{multirow}

\author{Remco Tuinier}
\email{r.tuinier@tue.nl}
\affiliation{Laboratory of Physical Chemistry, Department of Chemical Engineering and Chemistry, \& Institute for Complex Molecular Systems (ICMS), Eindhoven University of Technology, P.O. Box 513, 5600 MB, Eindhoven, the Netherlands}
\author{Anja Kuhnhold}
\email{anja.kuhnhold@physik.uni-freiburg.de}
\affiliation{Institute of Physics, University of Freiburg, Hermann-Herder-Str. 3, 79104 Freiburg, Germany}

\title%[An \textsf{achemso} demo]
{Equation of State of Charged Rod Dispersions}

\begin{document}

%\begin{tocentry}
%\includegraphics{figures/toc.pdf}
%\end{tocentry}

\begin{abstract}
  We study the accuracy of the theory of Stroobants, Lekkerkerker and Odijk [\textit{Macromolecules} \textbf{1986}, \textit{19}, 2232–2238], called SLO theory, to describe the thermodynamic properties of an isotropic fluid of charged rods. By incorporation of the effective diameter of the rods according to SLO theory into scaled particle theory (SPT) we obtain an expression for the rod concentration-dependent free volume fraction and the osmotic pressure of a collection of charged hard spherocylinders. The results are compared to Monte Carlo simulations. We find close agreement between the simulation results and the SLO-SPT predictions for not too large values of the Debye length and for high rod charge densities. The deviations increase with rod density, particularly at concentrations above which 
 isotropic--nematic phase transitions are expected.
\end{abstract}
\clearpage

\section{Introduction}
Dispersions of rod-like particles are of great interest because of the relatively large excluded volume between rods. If the length ($L$) to diameter ($D$) ratio is larger than about 4, see for instance Ref.~\citenum{Bolhuis1997}, hard rods can form liquid crystalline phases at low concentrations, as was already predicted by Onsager\cite{Onsager1949}. The use of liquid crystals in device applications such as displays has gained enormous interest over the last 50 years\cite{Drzaic1986,Schadt1996,Sheraw2002}, and responsive liquid crystals are a recent topic of development\cite{Haan2014,Bukusoglu2016,Vantomme2017}. Another reason for the increased interest into molecular and colloidal liquid crystals is the wealth of different possible phases with orientational order and/or partial positional order which appear\cite{
Thoen1984,Percec1995,Bolhuis1997,DogicFraden2001,Bakker2016,PetersPRE2020,Lopes2021}. Industrial applications are for instance the wet spinning of ultrastrong fibres\cite{Sawyer1986} and the increased interest in applying cellulose nanocrystals (CNC)\cite{Lu2010,Salas2014,Lagerwall2014,Honorato2018}. 
Understanding solutions of rod-like macromolecules is also of biological interest: think of the dense packing of rod-like DNA molecules in
virus heads\cite{ThompsonOdijk2004,Roos2007} or in the organization of cells\cite{Woldringh1999,Khmelinskaia2021}. Micro-organisms sometimes assume a spherocylinder shape, such as \textit{Escherichia coli}\cite{Osborn2007}, \textit{Campylobacter fetus}, \textit{Salmonella typhimurium}, and \textit{Mycobacterium tuberculosis} bacteria.
Assuming rod-like morphologies is considered to be a tool to gain a competitive advantage\cite{Young2007,Yang2016}.

Rod-like colloids often assume charges when they are dispersed in a polar solvent. CNCs are rod-like and are charged when dispersed in aqueous solution\cite{Eichhorn2010,Habibi2010}. Several types of viruses such as tobacco mosaic virus (TMV)\cite{Stupp2008,Fraden1989,Wen1989} or the filamentous bacteriophages \textit{fd}-virus\cite{Dogic1997,Grelet2014}, and its variants\cite{Janmey2014,Tarafder2020} like Pf1 and Pf4, are long and thin rods, assuming charged surfaces in water. In aqueous salt solutions charged rods are surrounded by double layers with an inhomogeneous distribution of co- and counterions\cite{Chapot2005}. The screening length of the extent of double layers is often described using the Debye length, while the magnitude of the inhomogeneity is determined by the surface charge density at the rod surface. Resulting double layer forces between charged colloidal rods in a polar solvent are specified by the range and the strength of the repulsive interaction, in a similar fashion as for colloidal spheres\cite{VerweyOverbeek1948}. Since the ionic strength affects the range of the double layer repulsion between the rods, it influences the rod concentration at which the dispersion undergoes phase transitions.     

Fraden and others\cite{Dogic1997,Adams1998,Adams1998a,Dogic2000,Dogic2001} showed that the salt concentration affects the isotropic--nematic phase transition of tobacco mosaic\cite{Fraden1989} and \textit{fd}\cite{Dogic2006b} viruses. 
It has also been shown that the phase transition concentration of CNC dispersions depends on the ionic strength.\cite{Honorato2016}

To theoretically predict such phase transitions, Onsager\cite{Onsager1949} proposed to describe charged rods as hard rods with an \textit{effective} diameter $D^\mathrm{eff} > D$, to mimic the hard core plus soft repulsion of charged rods. The Stroobants-Lekkerkerker-Odijk (SLO) approach\cite{SLO1986} nowadays is a standard method to estimate $D^\mathrm{eff}$. From this theory it follows that $D^\mathrm{eff}$ becomes larger when the Debye length and/or the surface charge are larger. 

Although the SLO theory is applied frequently its accuracy has, as far as we are aware, 
not been evaluated in detail. Vologodskii and Cozzarelli\cite{Vologodskii1995} tested different approaches for the electrostatic interactions in closed DNA chains modeled as chains of connected straight segments. They found a good agreement between the Debye-H\"uckel theory and the effective diameter model for the conformational properties of interest.
In this paper we verify the accuracy of SLO theory against computer simulations. We first show that determining the free volume fraction available directly provides the osmotic pressure, and thereby the equation of state. Subsequently, we analyze the accuracy of this approach. 

\clearpage

\section{Methods}

\subsection{Theory}\label{theory}

\subsubsection{Connection between $\alpha$ and the equation of state}\label{alphamuW}
The volume available in a system of interest is an important thermodynamic quantity. From Widom's particle insertion theorem\cite{Widom1963} it follows that the relative volume available for a particle upon insertion into a dispersion of particles, called the free volume fraction $\alpha$, is directly related to the immersion free energy $W$. The excess chemical potential of the particle $\mu_\mathrm{ex}=k_\mathrm{B}T\widetilde{\mu}_\mathrm{ex}=\mu-\mu{^{0}}-k_\mathrm{B}T \ln\phi $, where $\mu^{(0)}$ is the (reference) chemical potential and $\phi$ is the particle volume fraction, is related to $W$ through:
\begin{equation}\label{muexW}
\widetilde{\mu}_\mathrm{ex}=\frac{W}{k_\mathrm{B}T}=-\ln \alpha \text{.}
\end{equation}
Applying the Gibbs-Duhem relation:
\begin{equation}\label{GD}
\frac{\partial \widetilde{\Pi}}{\partial \phi}=\phi \frac{\partial \widetilde{\mu}}{\partial \phi} = 1+ \phi \frac{\partial \widetilde{\mu}_\mathrm{ex}}{\partial \phi}
 \text{,}
\end{equation}
to Eq.~\eqref{muexW} yields :
\begin{equation}\label{EOSalfa}
\widetilde{\Pi}=\phi[1-\ln \alpha(\phi)] + \int_{0}^{\phi} \ln \alpha(\phi') \mathrm{d}\phi'
 \text{,}
\end{equation}
which provides a direct relation\cite{Vortler2000} between the free volume fraction for a particle in a system and the osmotic pressure $\widetilde{\Pi}$, hence the equation of state. We apply this to the cases of hard rods and charged hard rods. Rods are described as spherocylinders, which are cylinders (with hard core volume $v_\text{r}$) with length $L$ capped by hemispheres with diameter $D$. It is assumed the spherocylinders only adopt isotropic configurations for the range of rod volume fractions studied here\cite{Bolhuis1997,PetersPRE2020}.

\subsubsection{Pressure of hard spherocylinders: SPT prediction}\label{theoryHSC}
We use scaled particle theory (SPT) to quantify the volume fraction $\phi_\text{r}$ dependence of the osmotic pressure ${\Pi}$ of a fluid of hard spherocylinders (HSC)\cite{Cotter1974}:
\begin{equation}
    \widetilde{\Pi} = 
   y + 
    \frac{3 \gamma (\gamma+1)}{3 \gamma - 1} \, y^{2} +\frac{12 \gamma^{\, 3}}{(3 \gamma -1)^{2}}\, y^{3} 
    \label{PiRods}
    \text{,}
\end{equation}
with $\widetilde{\Pi}={\Pi} v_\text{r}/{k_{\mathrm B}T}$ and $y={\phi_\text{r}}/({1-\phi_\text{r}})$. 
The dimensionless parameter $\gamma$ is related to the aspect ratio of the rods $L/D$ as $\gamma = 1 +{L}/{D}$. This result for the osmotic pressure accurately describes computer simulations of hard spherocylinders\cite{McGrother1996,Bolhuis1997}, see Ref.~\citenum{PetersPRE2020}. 
The free volume fraction $\alpha$ available to a HSC can be determined with SPT and reads\cite{Opdam2021b}:
\begin{subequations}   \label{alphaHSC}  
\begin{align}
\alpha &= (1-\phi_\text{r}) \exp \left[-\frac{6\gamma(1+\gamma)}{3 \gamma - 1}  \, y-\frac{18\gamma^3}{(3 \gamma - 1)^2}\, y^{2} - \widetilde{\Pi}\right] , \label{alphaHSCa} \\
&= (1-\phi_\text{r}) \exp \left[\frac{1-9\gamma-6\gamma^2}{3 \gamma - 1}  \, y-\frac{27 \gamma^3 + 6 \gamma^2-3\gamma}{(3 \gamma - 1)^2}\, y^{2} - \frac{12 \gamma^{\, 3}}{(3 \gamma -1)^{2}}\, y^{3} \right] ,\label{alphaHSCb} 
\end{align}
\end{subequations}
where the last expression results from insertion of Eq.~\eqref{PiRods} into Eq.~\eqref{alphaHSCa}.

\subsubsection{Pressure of charged spherocylinders}\label{SLOtheory}
\begin{figure}[tbp]
\centering
\includegraphics[trim=0 0 0 0,clip,width=0.3\columnwidth]{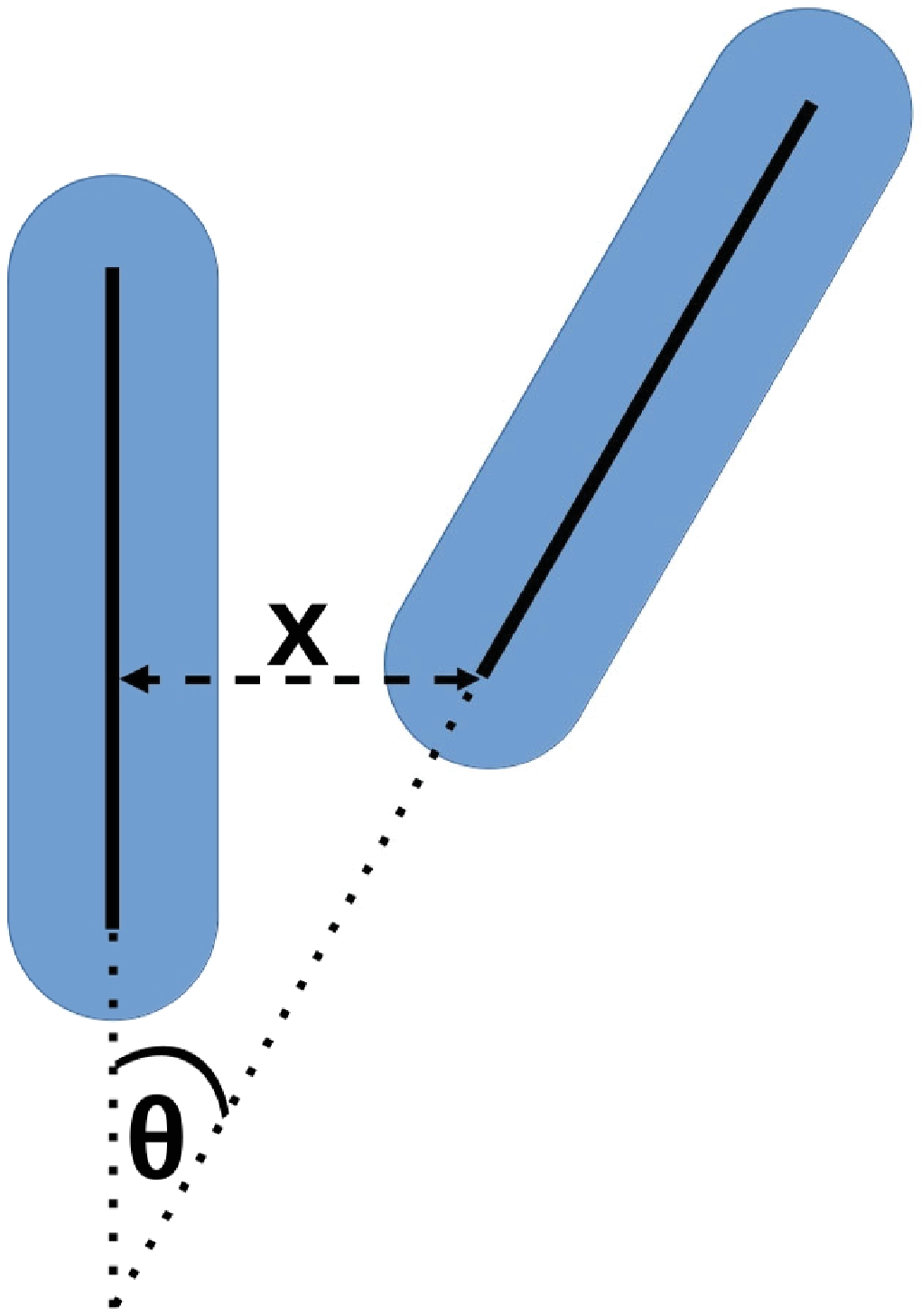}
\caption{Sketch to indicate the closest distance $x$ and the angle $\theta$ between two rods. The blue shape shows the hard spherocylinders (no overlap allowed) while the black solid lines mark their central axes which are decisive for the pair interaction Eq.~\eqref{rodsint}.}
\label{f:interaction}
\end{figure}

We now consider charged rods dispersed in a polar solvent that interact through a double layer repulsion next to the hard-core excluded volume interaction. This double layer interaction gives rise to an orientation-dependent soft repulsive interaction between the rods. The density of surface charge groups, which is directly related to the electrostatic surface potential $\Psi$ at the rod surface, determines the strength of the repulsion. The ionic strength of the medium dictates the Debye length, which mediates the range of the double layer repulsion\cite{Chapot2005}. Stroobants, Lekkerkerker and Odijk\cite{SLO1986} used the following expression for the pair interaction $U(x,\,\theta)$ between two charged rods:
\begin{equation} \label{rodsint}
\tilde{U}(x,\,\theta)=\frac{U(x,\,\theta)}{k_\mathrm{B}T}=\frac{A \mathrm{e}^{-\kappa x}}{\sin \theta} ,
\end{equation}
where $\kappa^{-1}$ is the ionic strength-dependent Debye screening length, $x$ is the closest distance between the center lines of the rods, and $\theta$ is the angle between the rods, cf. Fig.\ \ref{f:interaction}. 

\begin{figure}[tbp]
\centering
\includegraphics[trim=0 0 0 0,clip,width=0.5\columnwidth]{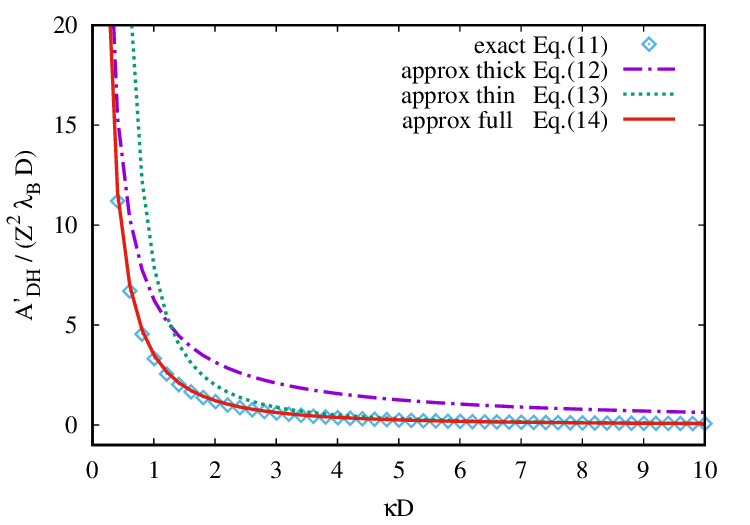}
\caption{Comparison of approximations for $A_{\mathrm{DH}}'$: exact expression (Eq.\ \ref{AprimeeqSLO}), proposed approximation (Eq.\ \ref{AprimeeqSLOapprox}), and the often used approximations for small and large $\kappa D$.}
\label{f:Aapprox}
\end{figure}

Incorporation of Eq.~\eqref{rodsint} into the Onsager free energy of infinitely long rods reveals that an effective rod diameter is given by:\cite{SLO1986}
\begin{equation} \label{SLOeq}
\frac{D^\mathrm{eff}}{D}=1+ \frac{\ln A'+ k_\mathrm{E}+\ln 2 - \frac{1}{2}}{\kappa D},
\end{equation}
where $k_\mathrm{E}$ is Euler's constant $\approx 0.5772156649$ and $A'=A \mathrm{e}^{-\kappa D}$ follows from:
\begin{equation} \label{Aprimeeq}
A'=\frac{\pi \zeta^2 \exp{[-\kappa D]}}{2 \kappa \lambda_\mathrm{B}},
\end{equation}
with $\lambda_\mathrm{B}$ the Bjerrum length. The parameter $\zeta$ is the proportionality constant of the outer part of the double layer electrostatic potential profile near a charged rod\cite{PhilipWooding1970}:
\begin{equation} \label{Woodingeq}
\frac{e \Psi(r)}{k_\mathrm{B}T} \sim \zeta K_0 (\kappa r) ,
\end{equation}
with $e$ the elementary charge, $r$ the distance from the center line of the rod and $K_0$ is the modified second kind of order 0 Bessel function. For a weakly charged rod the Debye--H{\"u}ckel approximation provides\cite{Hill1955}: 
\begin{equation} \label{Hilleq}
\frac{e \Psi_\mathrm{DH}(r)}{k_\mathrm{B}T} = \frac{4 Z  \lambda_\mathrm{B} K_0 (\kappa r)}{\kappa D K_1(\kappa D/2)} ,
\end{equation}
where $Z$ is the linear charge density (per unit length) of the rod and $K_1$ is the modified Bessel function of the second kind of order 1. Comparison of equations \eqref{Woodingeq} and \eqref{Hilleq} provides an expression for $\zeta$ which, after insertion into Eq.~\eqref{Aprimeeq}, gives:
\begin{equation} \label{AprimeeqSLO}
A_{\mathrm{DH}}'=\frac{8 \pi Z^2  \lambda_\mathrm{B} \exp{[-\kappa D]}}{\kappa^{3}D^2 {K_1}{^2}(\kappa D/2)}.
\end{equation}
For thick and thin double layers the following asymptotic analytical expressions have been derived\cite{Vroege1992}:
\begin{subequations}
\label{Aasymp}
\begin{align}
A_{\mathrm{DH}}' &\simeq \displaystyle{\frac{2 \pi Z^2  
\lambda_\mathrm{B}}{\kappa}} &  D  \ll \kappa^{-1}  \label{Athick}\\
   & \simeq \displaystyle{\frac{8Z^2 
\lambda_\mathrm{B}}{\kappa^2 D}}, &  
D \gg \kappa^{-1} \label{Athin}
\end{align}
\end{subequations}
We found that Eq.~\eqref{AprimeeqSLO} can be approximated accurately for arbitrary double layer thickness as:
\begin{equation} \label{AprimeeqSLOapprox}
A_{\mathrm{DH}}'^{\text{approx}}=\frac{8 Z^2  \lambda_\mathrm{B}/\kappa }{\kappa D + {4}/{\pi}}.
\end{equation}
The accuracy of this approximation is visualized in Fig.\ \ref{f:Aapprox}. The relative difference between the approximation and the exact expression is below 5\% for the whole range of $\kappa D$, cf.\ Fig.~S1 %\ref{SIe} 
in the SI.

The expression for the osmotic pressure in a system of charged rods is found by insertion of $D^\mathrm{eff}$ for $D$ into Eq.~\ref{PiRods}: 
\begin{equation}
    \widetilde{\Pi} = 
   y^\mathrm{eff} + 
    \frac{3 \gamma^{*} \left(\gamma^{*}+1\right)}{3 \gamma^{*} - 1} \, \left(y^\mathrm{eff}\right)^{2} +\frac{12 (\gamma^{*})^{3}}{(3 \gamma^{*} -1)^{2}}\, \left(y^\mathrm{eff}\right)^{3} 
    \label{PiChRods}
    \text{,}
\end{equation}
with $\gamma^{*} 
= 1 + {L}/{D^\mathrm{eff}}$
and $y^\mathrm{eff}={\phi^\mathrm{eff}_\text{r}}/({1-\phi^\mathrm{eff}_\text{r}})$, where $\phi_\mathrm{r}^\mathrm{eff}$ is the effective volume fraction of rods with diameter $D^\mathrm{eff}$. 

\subsection{Computer simulations}\label{sims}

In computer simulations, one cannot easily deal with infinitely ranged interactions, and, thus, has to define a maximum range, up to which pair interactions are taken into account. To not introduce further spurious effects, the interaction potential is also shifted so that it is zero when the distance between particles equals the defined range. This results in the following adaptation of the interaction potential $\tilde{U}_\mathrm{c}(x,\,\theta)={U_\mathrm{c}(x,\,\theta)}/{k_\mathrm{B}T}$ of Eq.\ (\ref{rodsint})
\begin{equation} \label{rodsintcutsh}
\tilde{U}_\mathrm{c}(x,\,\theta)=\begin{cases}
\infty & \text{if } x<D\quad \text{(hard core)}\\
\tilde{U}(x,\,\theta)-\tilde{U}(d_\mathrm{c},\,\theta)=\frac{A}{\sin \theta} \left(\mathrm{e}^{-\kappa x}-\mathrm{e}^{-\kappa d_\mathrm{c}}\right)&\text{if } D\le x<d_\mathrm{c} \\ 0 & \text{otherwise}\; ,
\end{cases}
\end{equation}
where $d_\mathrm{c}$ is the cutoff length.

To estimate the effective free volume fraction $\alpha^\mathrm{eff}$ available for the rods and the immersion free energy $W$, we apply test particle insertions. To test the relation between $\alpha^\mathrm{eff}$ and $W$ given in Eq.\ (\ref{muexW}), we compute $\alpha^{\rm ref}:=\langle \mathrm{e}^{-\Delta \tilde U_\mathrm{c}}\rangle$, where the brackets denote an average over $n_\mathrm{t}$ trial insertions and $n_\mathrm{c}$ independent configurations. $\Delta \tilde U_\mathrm{c}$ is the system's energy difference due to the particle insertion and an insertion is done by randomly choosing a position and an orientation for the inserted rod (which is removed again after the trial). For the effective free volume fraction $\alpha^\mathrm{eff}$, the relative number of successful trial insertions, $n_\mathrm{s}/n_\mathrm{t}$, is computed, where successful means that there is no overlap between the inserted rod and any other rod in the system assuming that the rods have diameter $D^\mathrm{eff}$. The quantity $\alpha^\mathrm{eff}$ is then approximated by the average $\langle n_\mathrm{s}/n_\mathrm{t}\rangle_{n_\mathrm{c}}$.

The systems contain $N_\mathrm{r}=1352$ rods and have periodic boundary conditions in all three dimensions. For each set $\left( A,\kappa,d_\mathrm{c}\right)$, we initialize the system at a density high enough to have a very small effective free volume fraction and with all rods pointing in the same direction. The system then relaxes to its equilibrium state, in which subsequently the measurements are done. The simulation boxes of equilibrated states are expanded to reduce the rod volume fraction, and, thus, get the density dependence of the observables (after re-equilibrating at the new density). The simulation method is Metropolis-Monte-Carlo, where single particle translations and rotations are used as trial moves, which are accepted with probability $\min\left(1, \mathrm{e}^{-\Delta\tilde U_\mathrm{c}} \right)$ with the energy difference before and after the move, $\Delta\tilde U_\mathrm{c}$. The acceptance rate for the trial moves is adjusted to be between $0.3$ and $0.4$ by changing the maximum displacement and rotation (unless the density is too low, which results in overall high acceptance rates).

\clearpage
\section{Results \& Discussion}

In this section we first show the accuracy of SPT to predict the free volume fraction and osmotic pressure using Eqs.\ (\ref{EOSalfa}),
(\ref{PiRods}) and (\ref{alphaHSC}) for hard spherocylinders by comparing the results with computer simulations. Since the simulations can only be performed by defining a cut-off length of the range of the soft interaction of Eq.\ (\ref{SLOeq}) we next study its proper value. Subsequently, we compare Monte Carlo simulation results for the free volume fractions of a dispersion of rods with an effective diameter against that of rods with a hard core but soft repulsion following Eq.\ (\ref{SLOeq}). Finally, we compare the theoretical predictions of the equation of state from theory (Eq.~(\ref{PiChRods})) with computer simulations.

\subsection{Bench-marking the approach for hard spherocylinders}
\begin{figure*}[tbp]
\centering
\begin{picture}(470,155)%(0.99\columnwidth,0.5\textheight)
\put(10,0){\includegraphics[trim=0 0 0 0,clip,width=0.5\columnwidth]{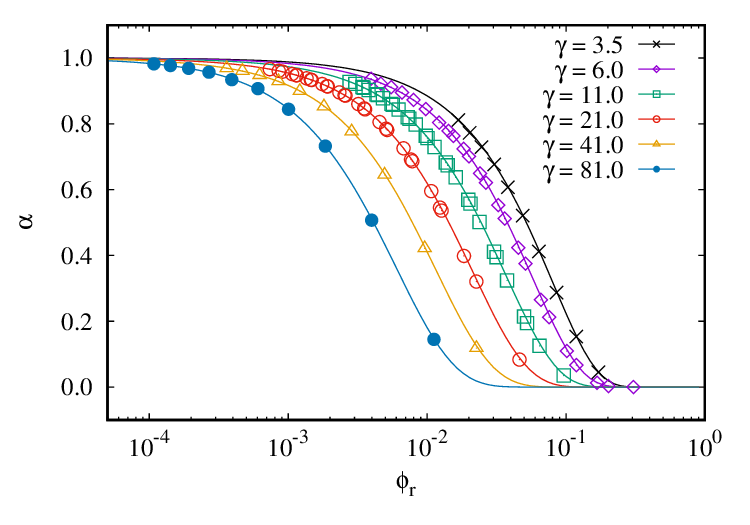}}
\put(240,0){\includegraphics[trim=0 0 0 0,clip,width=0.5\columnwidth]{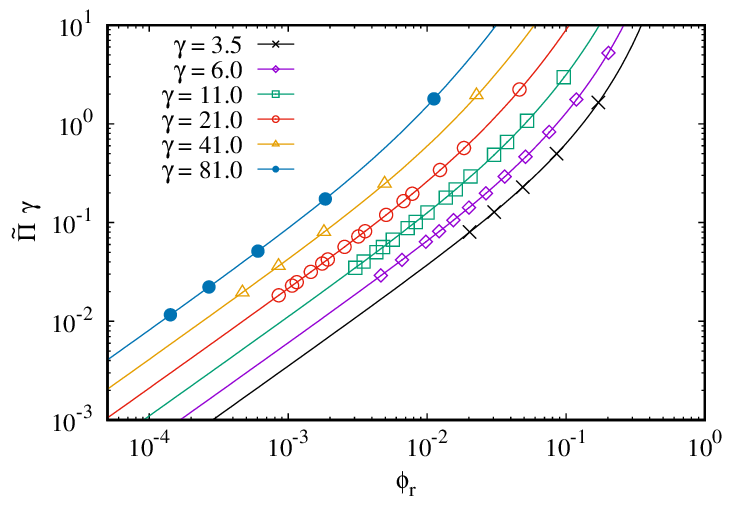}}
\put(10,144){\small{\textbf{(a)}}}
\put(240,144){\small{\textbf{(b)}}}
\end{picture}
\caption{Theoretical SPT predictions (curves) of (a) Eq.~(\ref{alphaHSC}) and (b) Eq.~(\ref{PiRods}) compared to simulation results (symbols) of the free volume fraction (a) and the osmotic pressure (b) of (uncharged) hard spherocylinders vs.\ rod volume fraction for different rod aspect ratios $\gamma=1+L/D$ as indicated in the legend.}
\label{f:alpha_theo_sim_HSC}
\end{figure*}

First we evaluate the accuracy of our method for hard spherocylinders for various values of $\gamma$. In a previous study,\cite{Opdam2022} we already confirmed the accuracy of Eq.\ (\ref{alphaHSC}) for a specific aspect ratio of $\gamma=6$. In Fig.~\ref{f:alpha_theo_sim_HSC}(a) the data points are Monte Carlo computer simulation results for the free volume fraction and the curves are predictions of Eq.\ (\ref{alphaHSC}). These MC simulation data points were used to compute the osmotic pressure using the exact relation Eq.~(\ref{EOSalfa}) and those results are the data in Fig.~\ref{f:alpha_theo_sim_HSC}(b). It follows that SPT correctly predicts the free volume fraction (Eq.\ (\ref{alphaHSC})) and osmotic pressure (Eq.\ (\ref{PiRods})) of hard spherocylinders in the isotropic phase state for a wide range of aspect ratios, cf.\ Figs\ \ref{f:alpha_theo_sim_HSC}(a) and (b).

\subsection{Effective diameter for a finite ranged interaction}
Eq.\ (\ref{SLOeq}) is valid for $A'\gtrsim 2$ and $d_\mathrm{c}\to \infty$, see Ref.~\citenum{SLO1986}. The exact result for the effective diameter and an arbitrary cutoff length $d_{\mathrm c}$ can be found by (numerically) evaluating the following integral:\cite{Onsager1949,SLO1986}
\begin{equation} \label{exactDeff}
\frac{D^\mathrm{eff}}{D}=1+\frac{4}{\pi}\int\int \tilde{B}_{2,\text{el}}(\sin\theta) f(\Omega) f(\Omega') \mathrm{d}\Omega \mathrm{d}\Omega'\;,
\end{equation}
with the orientation distribution function in the isotropic state, $f(\Omega)=1/(4\pi)$,
and the scaled electrostatic excluded volume
\begin{align}
 \tilde{B}_{2,\text{el}}(\sin\theta) &= \frac{\sin\theta}{D} \int_D^{d_\mathrm{c}}\left( 1-\mathrm{e}^{-\tilde{U}_\mathrm{c}(x,\,\theta)}\right)\mathrm{d}x \label{elexvol1}\\
 &= \frac{\sin\theta}{\kappa D} \left(\kappa\left(d_\mathrm{c} - D\right) + 
   \mathrm{e}^{\tilde{U}(d_\mathrm{c},\,\theta)} \left[\mathrm{Ei}\left(-\tilde{U}(d_\mathrm{c},\,\theta)\right) - 
      \mathrm{Ei}\left(-\tilde{U}(D,\,\theta)\right)\right]\right) \label{elexvol2}\;,
\end{align}
%Note: in the second line it is $U$ and not $U_c$.
where the exponential integral is defined through  %$\mathrm{Ei}(z)=-\int_{-z}^\infty \mathrm{e}^{-t}/t ~\mathrm{d}t$.
$\mathrm{Ei}(z)=\int_{-\infty}^z \mathrm{e}^{t}/t ~\mathrm{d}t$.

For $d_\mathrm{c}\to \infty$, Eq.\ (\ref{elexvol2}) becomes
\begin{align}
 \tilde{B}_{2,\text{el}}(\sin\theta) &= \frac{\sin\theta}{\kappa D} 
    \left[\ln\left(\tilde{U}(D,\,\theta)\right) + \gamma - \mathrm{Ei}\left(-\tilde{U}(D,\,\theta)\right)\right] \label{elexvolinf}\;.
\end{align}
In this case, the relative difference between the approximation, Eq.\ (\ref{SLOeq}), and the exact value of the effective diameter $D^\mathrm{eff}$, Eq.\ (\ref{exactDeff}) is less than 2.2\% for $A'\ge 2$. For a finite cutoff length, one can determine the value of $d_\mathrm{c}$, for which the relative difference between the effective diameter using the approximation of an infinite interaction range and the exact effective diameter is also less than 2.2\%. This threshold cutoff length, $d_\mathrm{c}^\mathrm{t}$, decreases with $\kappa$ and increases with $A$. We provide some exemplary values for $d_\mathrm{c}^\mathrm{t}/D$ in Table \ref{tabdct}.

\begin{table}
    \caption{{Examples for the threshold of the cutoff length, $d_\mathrm{c}^\mathrm{t}/D$, beyond which the relative difference between the exact result and the infinite range approximation for the effective diameter is smaller than 2.2\%. The corresponding effective diameter, $D^\mathrm{eff}/D$, is given in square brackets.}}
\begin{center}
    \begin{tabular*}{\linewidth}{c @{\extracolsep{\fill}} cccc}
    \hline
    \hline
     \diagbox{$\kappa D$}{\raisebox{0.5ex}{$A$}} & 1 & 4 & 16
     \\ \hline
    0.5 & 11.3 [2.2] & 12.6 [4.3] & 14.2 [6.9] \\
    1.0 & 5.3 [1.4] & 6.3 [2.2] & 7.1 [3.5] \\ 
    2.0 & 2.3 [1.1] & 3.0 [1.3] & 3.5 [1.7] \\ \hline\hline
    \end{tabular*}
\end{center}
    \label{tabdct}
\end{table}

Thus, if $d_\mathrm{c}>d_\mathrm{c}^\mathrm{t}$ and $A' \ge 2$, Eq.\ (\ref{SLOeq}) gives a very close approximation for the effective diameter; if $d_\mathrm{c}>d_\mathrm{c}^\mathrm{t}$ and $A' < 2$, $D^\mathrm{eff}$ can be approximated by using Eq.\ (\ref{elexvolinf}) for evaluating Eq.\ (\ref{exactDeff}); and if $d_\mathrm{c}\le d_\mathrm{c}^\mathrm{t}$, the exact value from using Eq.\ (\ref{elexvol2}) should be applied.

\subsection{Comparison of $\langle \mathrm{e}^{-\tilde{U}_\mathrm{c}}\rangle$ and the effective free volume fraction}

\begin{figure*}[tbp]
\centering
\begin{picture}(470,310)%(0.99\columnwidth,0.5\textheight)
\put(10,155){\includegraphics[trim=0 0 0 0,clip,width=0.5\columnwidth]{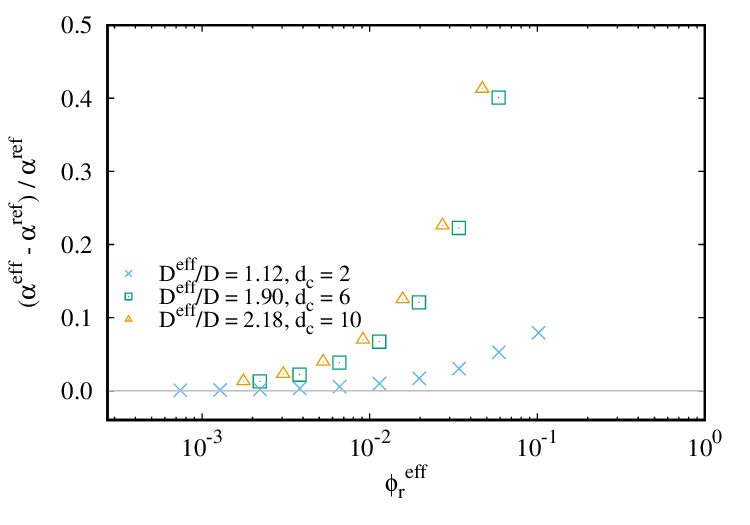}}
\put(50,242){\includegraphics[trim=0 0 0 0,clip,width=0.2\columnwidth]{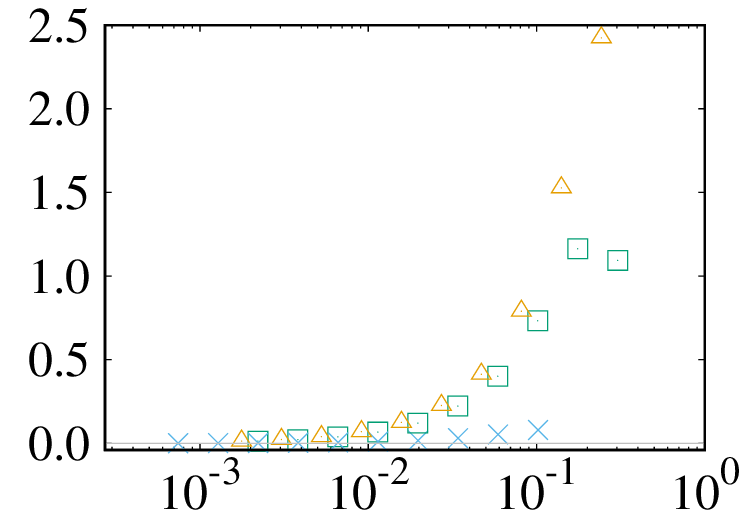}}
\put(240,155){\includegraphics[trim=0 0 0 0,clip,width=0.5\columnwidth]{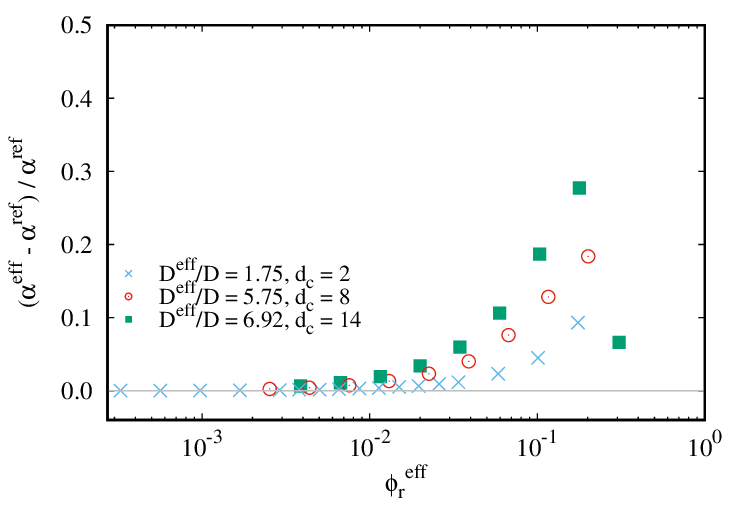}}
\put(10,0){\includegraphics[trim=0 0 0 0,clip,width=0.5\columnwidth]{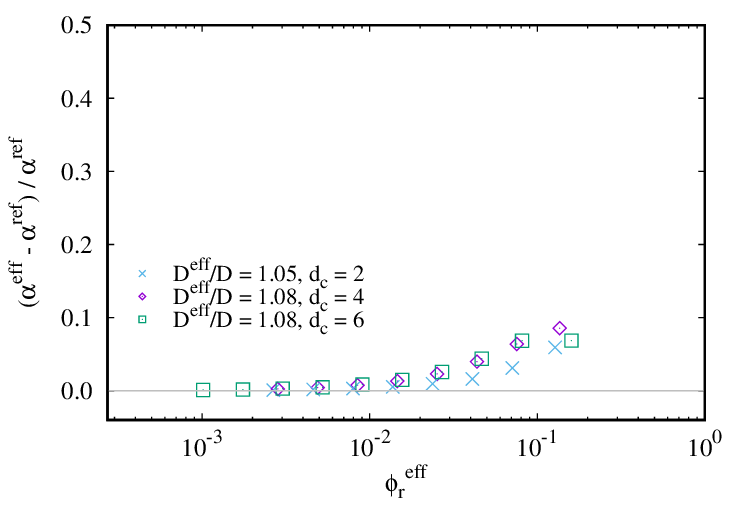}}
\put(240,0){\includegraphics[trim=0 0 0 0,clip,width=0.5\columnwidth]{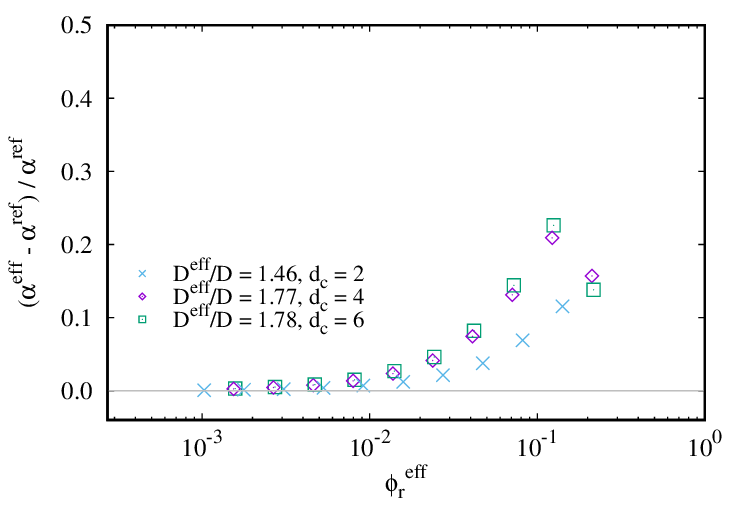}}
\put(10,298){\small{\textbf{(a)}}}
\put(240,298){\small{\textbf{(b)}}}
\put(10,144){\small{\textbf{(c)}}}
\put(240,144){\small{\textbf{(d)}}}
\end{picture}
\caption{Relative difference between the effective free volume fraction and the reference from the immersion free energy vs.\ effective volume fraction for different cutoff lengths as indicated in the legend and $(A,\kappa D)=$ $(1.0,0.5)$(a), $(16.0,0.5)$(b), $(1.0,2.0)$(c), $(16.0,2.0)$(d). The hard rod aspect ratio is $L/D+1=11$ and the effective diameter ratio $D^\mathrm{eff}/D$
%aspect ratio $\gamma=L/D^\mathrm{eff}+1$ 
is indicated in the legend. The inset in (a) shows the whole range of relative differences while the main figures show the same range for (a)-(d). (Note: We only show data with at least 10 successful insertions during the simulation run. The decreasing difference seen for some of the largest rod volume fractions is due to both free volume fraction estimates getting close to zero.)}
\label{f:rel_alpha_omega_10}
\end{figure*}
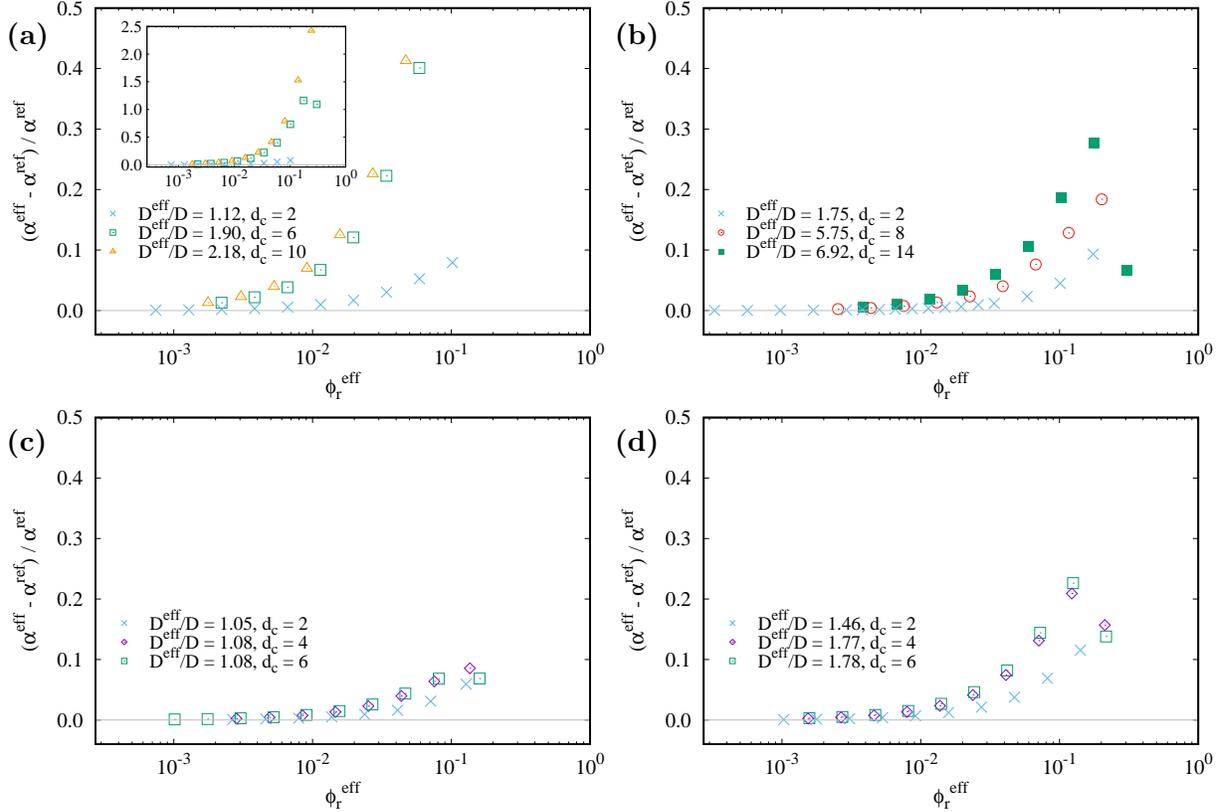

The effective free volume fraction is less expensive to compute compared to the true immersion free energy. Thus, it would be a great advantage, if the $D^\mathrm{eff}$ approach gives a good approximation of the immersion free energy. Figure \ref{f:rel_alpha_omega_10} (and Figs.~S2 and S3 
%\ \ref{SIa} and \ref{SIc} 
in the SI) shows the relative difference between $\alpha^\mathrm{eff}$ and $\alpha^{\rm ref}$ for different sets $(A,\kappa D)$ and different cutoff lengths versus the effective volume fraction of rods, $\phi_\mathrm{r}^\mathrm{eff}=N_{\mathrm r} \pi (L/4 + D^\mathrm{eff}/6) (D^\mathrm{eff})^2/V$. The difference varies a lot depending on the specific set of parameters. It increases with the rod density and with the cutoff length which can be explained as follows: The rods in the system can have a smaller shortest distance of their axes than the effective diameter, whereas the inserted rod only fits when its shortest distance to all other rods is at least $D^\mathrm{eff}$. That means if the rods truly had a hard diameter of $D^\mathrm{eff}$, the respective free volume would be smaller. This effect increases with the density because at higher density, more rods have a shortest distance between $D$ and $D^\mathrm{eff}$. And the effect increases when the effective diameter is larger, which is the case for increasing $d_\mathrm{c}$. Because the estimation of the effective diameter stems from a second virial approximation, it is not too surprising that the accuracy breaks down for denser systems. 

We have assumed the effective rod diameter is independent of the charged rod density. When it comes to studying the isotropic-nematic transition and other phase equilibria at higher rod concentrations this may lead to deviations. In practice, the assumption that Eq.~(\ref{rodsint}) is independent of the rod concentration will become inaccurate as the charged rods themselves will also contribute to the effective value of the screening length $\kappa^{-1}$. In addition, also the twisting effect\cite{SLO1986,Odijk1986,Drwenski2016} needs to be taken into account as soon as the rods undergo phase transitions. To improve the $D^\mathrm{eff}$ approach one could, for example, follow the avoidance model for soft particles that results in a reduction of the effective diameter with increasing concentration of the particles.\cite{Han1996, Kramer1999}

\begin{figure*}[tbp]
\centering
\begin{picture}(470,310)%(0.99\columnwidth,0.5\textheight)
\put(10,155){\includegraphics[trim=0 0 0 0,clip,width=0.5\columnwidth]{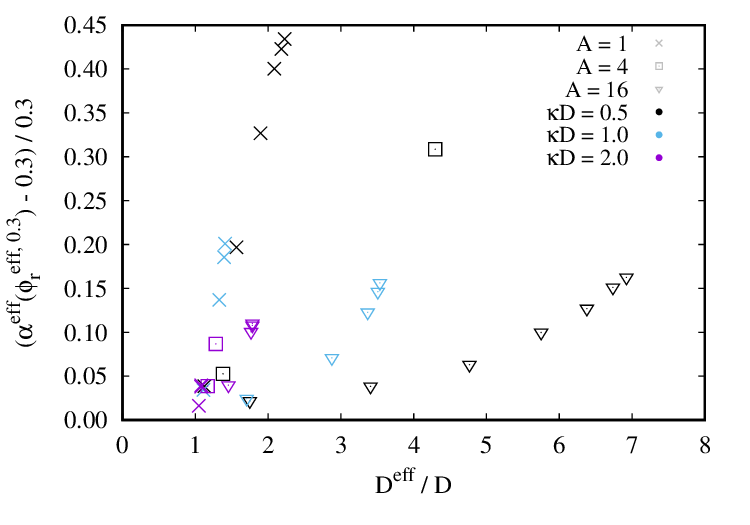}}
\put(240,155){\includegraphics[trim=0 0 0 0,clip,width=0.5\columnwidth]{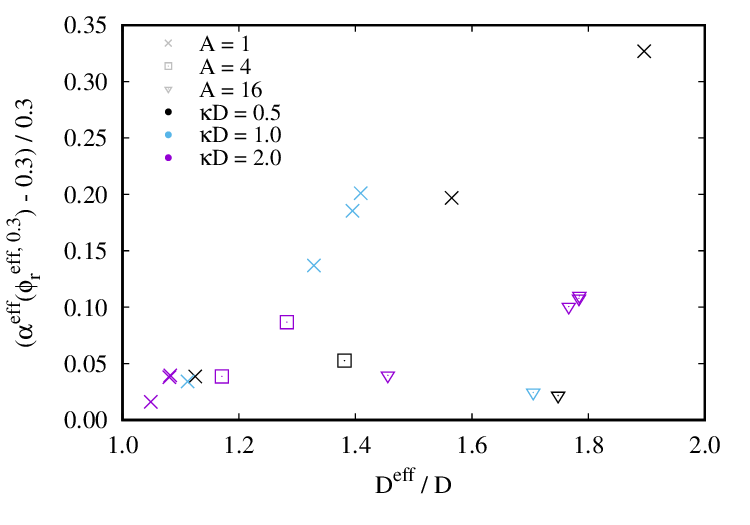}}
\put(10,0){\includegraphics[trim=-28 0 0 0,clip,width=0.5\columnwidth]{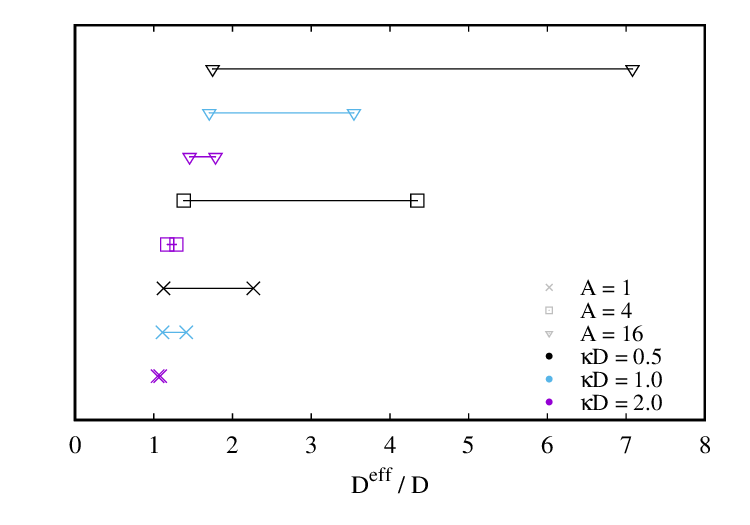}}
\put(240,0){\includegraphics[trim=-28 0 0 0,clip,width=0.5\columnwidth]{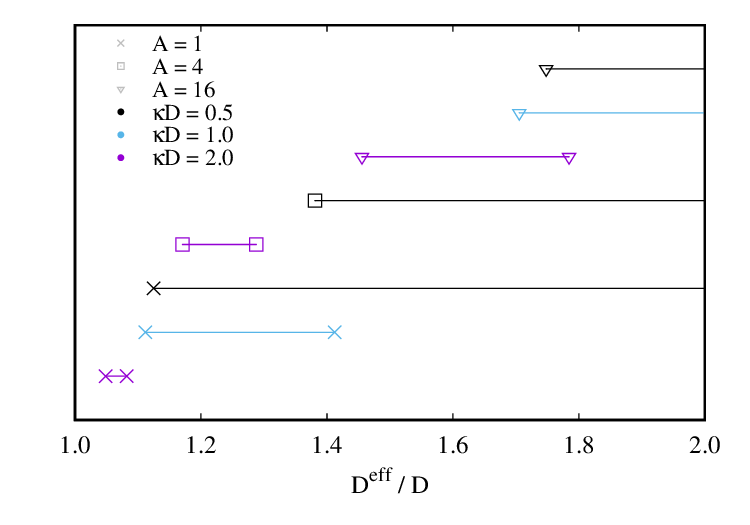}}
\put(10,298){\small{\textbf{(a)}}}
\put(240,298){\small{\textbf{(b)}}}
\put(10,144){\small{\textbf{(c)}}}
\put(240,144){\small{\textbf{(d)}}}
\end{picture}
\caption{(a)-(b) Relative difference between the effective free volume fraction and the reference from the immersion free energy at the effective volume fraction for which the reference equals 0.3 vs.\ effective diameter for different $A$ and $\kappa D$. Identical symbols refer to the same $(A,\kappa D)$ but different cutoff lengths (resulting in different effective diameter ratios). The hard rod aspect ratio is $L/D+1=11$. (a) Full range of effective diameters. (b) Small effective diameters. (c)-(d) Accessible range of effective diameter ratios for given $(A,\kappa D)$ and varying $d_\mathrm{c}\ge 2$.}
\label{f:rel_alpha_0.3_10}
\end{figure*}

In order to further quantify the accuracy of the $D^\mathrm{eff}$ approach for different combinations of $(A, \kappa D, d_\mathrm{c})$, we do the following. (i) We interpolate the data for $\alpha^\mathrm{eff}(\phi_\mathrm{r}^\mathrm{eff})$ and $\alpha^{\rm ref}(\phi_\mathrm{r}^\mathrm{eff})$. (ii) We find the effective rod volume fraction ($\phi_\mathrm{r}^\mathrm{eff, 0.3}$) for which the reference free volume fraction value equals $\alpha^{\rm ref} =$ 0.3. (iii) We compute the relative difference between the effective free volume fraction at this rod volume fraction and the reference free volume fraction value of 0.3\footnote{One can also use other reference values; the results are qualitatively the same.}. The results are shown in Fig.\ \ref{f:rel_alpha_0.3_10} (and in Figs.~S4 and S5 
%~\ref{SIb} and \ref{SId} 
in the SI). 

Figures \ref{f:rel_alpha_0.3_10}(a) and (b) provide an overview of the deviations of the $D^\mathrm{eff}$ approach from the results obtained using soft repulsive rods. It also identifies (Figs.\ \ref{f:rel_alpha_0.3_10}(c) and (d) ) the range of $D^\mathrm{eff}/D$ values over which the $D^\mathrm{eff}$ can be accurately applied for some illustrative ($A$, $\kappa D$) values.

\subsection{Comparison of theoretical predictions and simulations}

\begin{figure*}[tbp]
\centering
\begin{picture}(470,310)%(0.99\columnwidth,0.5\textheight)
\put(10,155){\includegraphics[trim=0 0 0 0,clip,width=0.5\columnwidth]{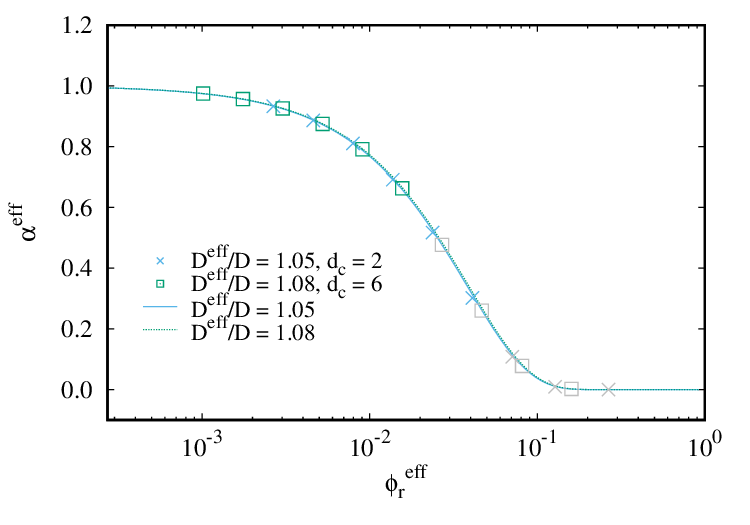}}
\put(240,155){\includegraphics[trim=0 0 0 0,clip,width=0.5\columnwidth]{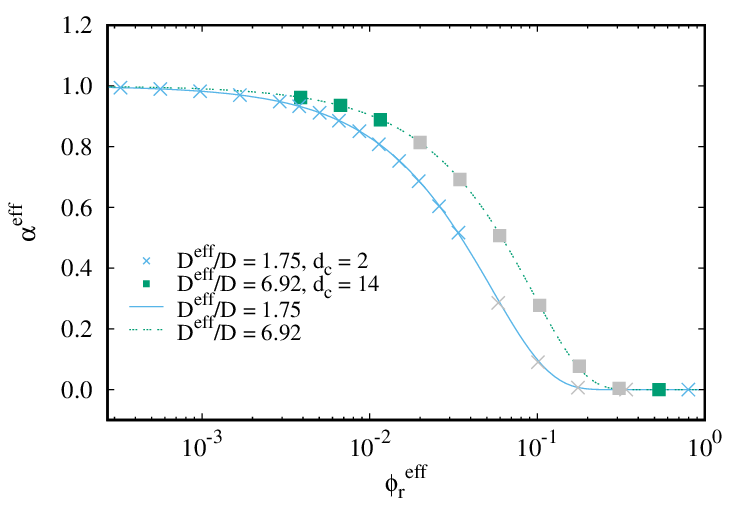}}
\put(10,0){\includegraphics[trim=0 0 0 0,clip,width=0.5\columnwidth]{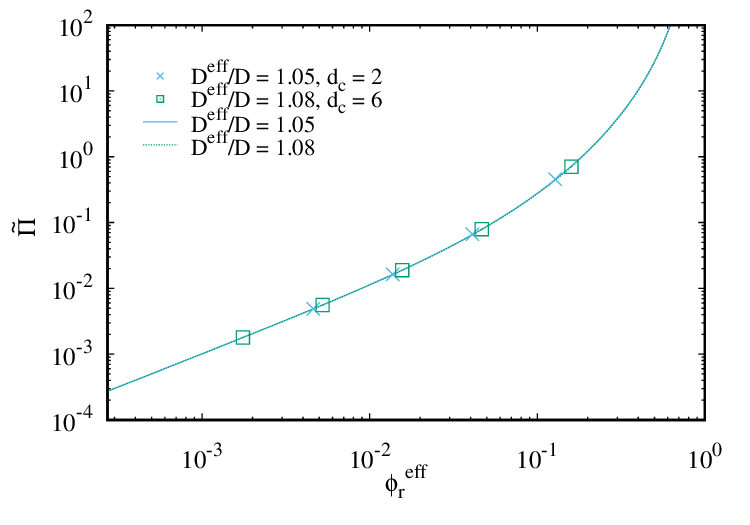}}
\put(240,0){\includegraphics[trim=0 0 0 0,clip,width=0.5\columnwidth]{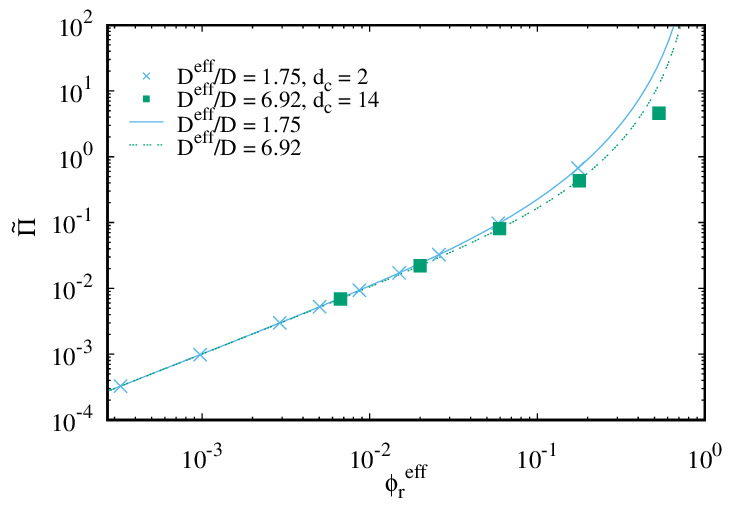}}
\put(10,299){\small{\textbf{(a)}}}
\put(240,299){\small{\textbf{(b)}}}
\put(10,144){\small{\textbf{(c)}}}
\put(240,144){\small{\textbf{(d)}}}
\end{picture}
\caption{Theoretical prediction and simulation result of the effective free volume fraction (a,b) and the osmotic pressure (c,d) vs.\ effective rod volume fraction for different cutoff lengths as indicated in the legend and $(A,\kappa D)=$ $(1.0,2.0)$(a,c), $(16.0,0.5)$(b,d). The hard rod aspect ratio is $L/D+1=11$ and the effective diameter ratio $D^\mathrm{eff}/D$
%aspect ratio $\gamma=L/D^\mathrm{eff}+1$ 
is indicated in the legend. Points in (a,b) are shown in grey when the relative difference between the effective free volume fraction and the reference from the immersion free energy is larger than 2.2\%, i.e.\ when there already is a systematic error in using the effective free volume to study thermodynamic properties.}
\label{f:alpha_theo_sim}
\end{figure*}

\begin{figure*}[tbp]
\centering
\begin{picture}(470,310)%(0.99\columnwidth,0.5\textheight)
\put(10,155){\includegraphics[trim=0 0 0 0,clip,width=0.5\columnwidth]{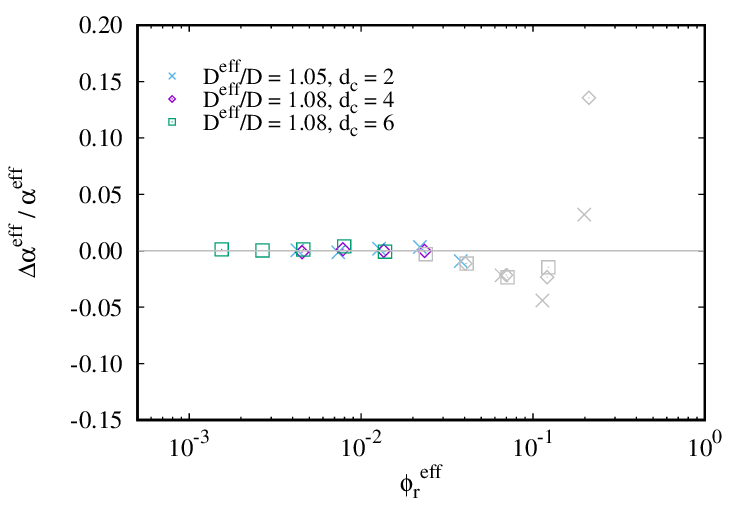}}
\put(240,155){\includegraphics[trim=0 0 0 0,clip,width=0.5\columnwidth]{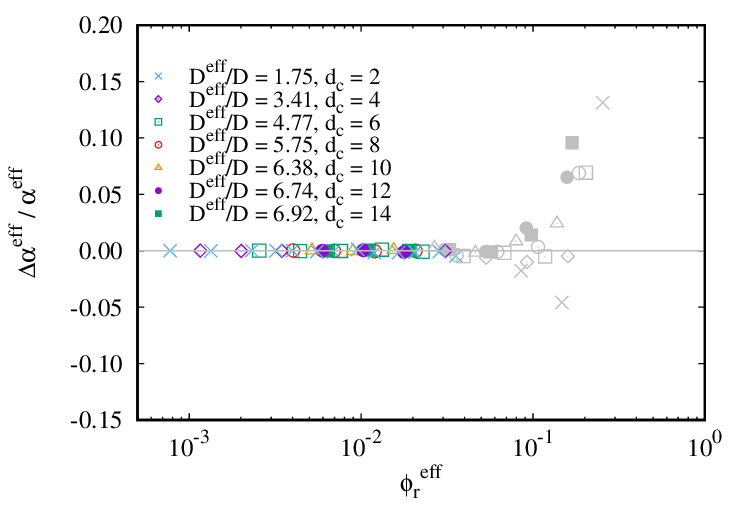}}
\put(10,0){\includegraphics[trim=0 0 0 0,clip,width=0.5\columnwidth]{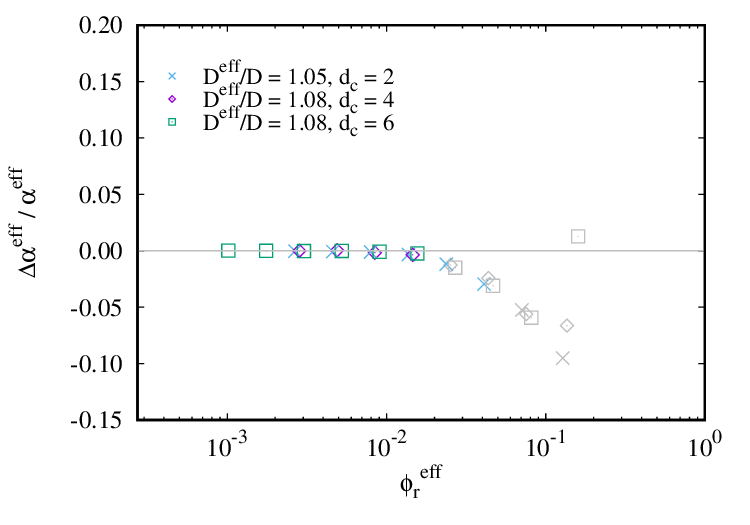}}
\put(240,0){\includegraphics[trim=0 0 0 0,clip,width=0.5\columnwidth]{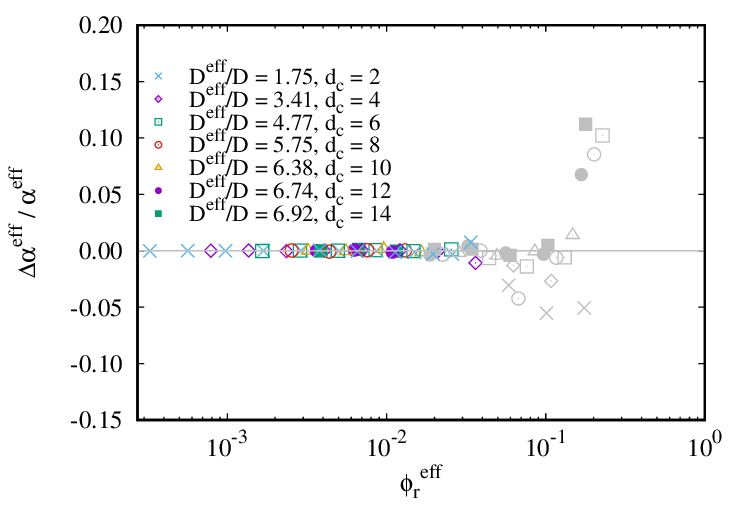}}
\put(10,298){\small{\textbf{(a)}}}
\put(240,298){\small{\textbf{(b)}}}
\put(10,144){\small{\textbf{(c)}}}
\put(240,144){\small{\textbf{(d)}}}
\end{picture}
\caption{Relative difference between the theoretical and simulation value of the effective free volume fraction vs.\ effective rod volume fraction for different cutoff lengths as indicated in the legend and $(A,\kappa D,L/D)=$ $(1.0,2.0,5)$(a), $(16.0,0.5,5)$(b), $(1.0,2.0,10)$(c), $(16.0,0.5,10)$(d). The effective diameter ratio $D^\mathrm{eff}/D$ is indicated in the legend. Points are shown in gray when the relative difference between the effective free volume fraction and the reference from the immersion free energy is larger than 2.2\%, i.e.\ when there already is a systematic error in using the effective free volume to study thermodynamic properties.}
\label{f:rel_alpha_theo_sim}
\end{figure*}
In Fig.\ \ref{f:alpha_theo_sim} we directly compare the theoretical predictions (SLO + SPT) with the simulation results for the effective free volume fraction and the resulting osmotic pressure. The agreement is almost perfect for all studied aspect ratios, cutoff lengths, and interaction parameters. Only when the effective free volume fraction goes to very small values (i.e., high effective rod concentrations), the relative difference between simulation and theory starts to increase, cf.\ Fig.\ \ref{f:rel_alpha_theo_sim}.
The conclusion from Fig.\ \ref{f:rel_alpha_theo_sim} is that when the system volume fraction is small enough so that the effective free volume fraction gives a good estimate of the immersion free energy, then the theoretically predicted value for the effective free volume fraction is very close to the value obtained from simulations. The deviation between theory and simulation for larger volume fractions is again related to overlaps of the effective shells; the theory assumes no overlap whereas there can be overlaps in the simulated system so that the actual volume fractions differ.

\clearpage 

\section{Conclusions}
In this paper we have studied the equation of state of charged rods. A frequently used theory for the thermodynamic properties of charged rods is that by Odijk, Stroobants and Lekkerkerker\cite{SLO1986}. This SLO theory showed that the free energy for infinitely long rods interacting through a hard-core double layer potential can be described via an effective diameter of the rods. Here we have verified the accuracy of this approach for finite sized rods.

We first showed that scaled particle theory (SPT) provides accurate results for the concentration and aspect ratio dependence of rods in the isotropic phase by comparison of the free volume fraction and osmotic pressure of hard rods. Next, we extended this approach by incorporating SLO theory into SPT to predict the osmotic pressure for charged rods and compare the results with computer simulations.

It is found that the accuracy of using the effective diameter following SLO theory depends on the Debye length, charge density of the rods and the rod concentration. Deviations decrease when the rod charge density is higher and the Debye length is smaller. For denser rods the deviations increase. It is realized however that for dense rods phase transitions are expected to nematic, smectic or crystalline phases where rod-mediated screening will reduce the effective diameter. We have the ambition to explore these phase transitions of charged rods in subsequent work.

%\begin{suppinfo}
%The relative difference between
%the approximation and the exact expression for $A'$, and some computer simulation results for other rod aspect ratios as those presented in the main text are available in the Supporting Information file.
%\end{suppinfo}

\clearpage

\begin{acknowledgement}
This work was financially supported by the European Union's Horizon 2020 research and innovation program by the EIC Pathfinder Open “INTEGRATE”, grant agreement 101046333 and by the Dutch Ministry of Education, Culture and Science (Gravity Program 024.005.020 – Interactive Polymer Materials IPM). R.T. thanks Dr. Joeri Opdam, Eleonora Foschino (TU Eindhoven) and Dr. Vincent F. D. Peters (Utrecht University) for useful discussions. Vincent is also thanked for checking the validity of Eq.~\eqref{AprimeeqSLOapprox} at an early stage. 
\end{acknowledgement}
 
%\bibliography{refschargedrods}

\begin{mcitethebibliography}{59}
	\providecommand*\natexlab[1]{#1}
	\providecommand*\mciteSetBstSublistMode[1]{}
	\providecommand*\mciteSetBstMaxWidthForm[2]{}
	\providecommand*\mciteBstWouldAddEndPuncttrue
	{\def\EndOfBibitem{\unskip.}}
	\providecommand*\mciteBstWouldAddEndPunctfalse
	{\let\EndOfBibitem\relax}
	\providecommand*\mciteSetBstMidEndSepPunct[3]{}
	\providecommand*\mciteSetBstSublistLabelBeginEnd[3]{}
	\providecommand*\EndOfBibitem{}
	\mciteSetBstSublistMode{f}
	\mciteSetBstMaxWidthForm{subitem}{(\alph{mcitesubitemcount})}
	\mciteSetBstSublistLabelBeginEnd
	{\mcitemaxwidthsubitemform\space}
	{\relax}
	{\relax}
	
	\bibitem[Bolhuis and Frenkel(1997)Bolhuis, and Frenkel]{Bolhuis1997}
	Bolhuis,~P.; Frenkel,~D. Tracing the phase boundaries of hard spherocylinders. \emph{J. Chem. Phys.} \textbf{1997}, \emph{106}, 666--687\relax
	\mciteBstWouldAddEndPuncttrue
	\mciteSetBstMidEndSepPunct{\mcitedefaultmidpunct}
	{\mcitedefaultendpunct}{\mcitedefaultseppunct}\relax
	\EndOfBibitem
	\bibitem[Onsager(1949)]{Onsager1949}
	Onsager,~L. The effects of shape on the interaction of colloidal particles. \emph{Ann. NY. Acad. Sci.} \textbf{1949}, \emph{51}, 627--659\relax
	\mciteBstWouldAddEndPuncttrue
	\mciteSetBstMidEndSepPunct{\mcitedefaultmidpunct}
	{\mcitedefaultendpunct}{\mcitedefaultseppunct}\relax
	\EndOfBibitem
	\bibitem[Drzaic(1986)]{Drzaic1986}
	Drzaic,~P.~S. Polymer dispersed nematic liquid crystal for large area displays and light valves. \emph{J. Appl. Phys.} \textbf{1986}, \emph{60}, 2142--2148\relax
	\mciteBstWouldAddEndPuncttrue
	\mciteSetBstMidEndSepPunct{\mcitedefaultmidpunct}
	{\mcitedefaultendpunct}{\mcitedefaultseppunct}\relax
	\EndOfBibitem
	\bibitem[Schadt \latin{et~al.}(1996)Schadt, Seiberle, and Schuster]{Schadt1996}
	Schadt,~M.; Seiberle,~H.; Schuster,~A. Optical patterning of multi-domain liquid-crystal displays with wide viewing angles. \emph{Nature} \textbf{1996}, \emph{381}, 212–215\relax
	\mciteBstWouldAddEndPuncttrue
	\mciteSetBstMidEndSepPunct{\mcitedefaultmidpunct}
	{\mcitedefaultendpunct}{\mcitedefaultseppunct}\relax
	\EndOfBibitem
	\bibitem[Sheraw \latin{et~al.}(2002)Sheraw, Zhou, Huang, Gundlach, Jackson, Kane, Hill, Hammond, Campi, Greening, Francl, and West]{Sheraw2002}
	Sheraw,~C.~D.; Zhou,~L.; Huang,~J.~R.; Gundlach,~D.~J.; Jackson,~T.~N.; Kane,~M.~G.; Hill,~I.~G.; Hammond,~M.~S.; Campi,~J.; Greening,~B.~K. \latin{et~al.}  Organic thin-film transistor-driven polymer-dispersed liquid crystal displays on flexible polymeric substrates. \emph{Appl. Phys. Lett.} \textbf{2002}, \emph{80}, 1088--1090\relax
	\mciteBstWouldAddEndPuncttrue
	\mciteSetBstMidEndSepPunct{\mcitedefaultmidpunct}
	{\mcitedefaultendpunct}{\mcitedefaultseppunct}\relax
	\EndOfBibitem
	\bibitem[de~Haan \latin{et~al.}(2014)de~Haan, Verjans, Broer, Bastiaansen, and Schenning]{Haan2014}
	de~Haan,~L.~T.; Verjans,~J. M.~N.; Broer,~D.~J.; Bastiaansen,~C. W.~M.; Schenning,~A. P. H.~J. Humidity-responsive liquid crystalline polymer actuators with an asymmetry in the molecular trigger that bend, fold, and curl. \emph{J. Am. Chem. Soc.} \textbf{2014}, \emph{136}, 10585--10588\relax
	\mciteBstWouldAddEndPuncttrue
	\mciteSetBstMidEndSepPunct{\mcitedefaultmidpunct}
	{\mcitedefaultendpunct}{\mcitedefaultseppunct}\relax
	\EndOfBibitem
	\bibitem[Bukusoglu \latin{et~al.}(2016)Bukusoglu, Bedolla~Pantoja, Mushenheim, Wang, and Abbott]{Bukusoglu2016}
	Bukusoglu,~E.; Bedolla~Pantoja,~M.; Mushenheim,~P.~C.; Wang,~X.; Abbott,~N.~L. Design of responsive and active (soft) materials using liquid crystals. \emph{Ann. Rev. Chem. Biomol. Engin.} \textbf{2016}, \emph{7}, 163--196\relax
	\mciteBstWouldAddEndPuncttrue
	\mciteSetBstMidEndSepPunct{\mcitedefaultmidpunct}
	{\mcitedefaultendpunct}{\mcitedefaultseppunct}\relax
	\EndOfBibitem
	\bibitem[Vantomme \latin{et~al.}(2017)Vantomme, Gelebart, Broer, and Meijer]{Vantomme2017}
	Vantomme,~G.; Gelebart,~A.~H.; Broer,~D.~J.; Meijer,~E.~W. Preparation of liquid crystal networks for macroscopic oscillatory motion induced by light. \emph{J. Vis. Exp.} \textbf{2017}, \emph{e56266}\relax
	\mciteBstWouldAddEndPuncttrue
	\mciteSetBstMidEndSepPunct{\mcitedefaultmidpunct}
	{\mcitedefaultendpunct}{\mcitedefaultseppunct}\relax
	\EndOfBibitem
	\bibitem[Thoen \latin{et~al.}(1984)Thoen, Marynissen, and Van~Dael]{Thoen1984}
	Thoen,~J.; Marynissen,~H.; Van~Dael,~W. Nematic-smectic-A tricritical point in alkylcyanobiphenyl liquid crystals. \emph{Phys. Rev. Lett.} \textbf{1984}, \emph{52}, 204--207\relax
	\mciteBstWouldAddEndPuncttrue
	\mciteSetBstMidEndSepPunct{\mcitedefaultmidpunct}
	{\mcitedefaultendpunct}{\mcitedefaultseppunct}\relax
	\EndOfBibitem
	\bibitem[Percec \latin{et~al.}(1995)Percec, Chu, Ungar, and Zhou]{Percec1995}
	Percec,~V.; Chu,~P.; Ungar,~G.; Zhou,~J. Rational design of the first nonspherical dendrimer which displays calamitic nematic and smectic thermotropic liquid crystalline phases. \emph{J. Am. Chem. Soc.} \textbf{1995}, \emph{117}, 11441--11454\relax
	\mciteBstWouldAddEndPuncttrue
	\mciteSetBstMidEndSepPunct{\mcitedefaultmidpunct}
	{\mcitedefaultendpunct}{\mcitedefaultseppunct}\relax
	\EndOfBibitem
	\bibitem[Dogic and Fraden(2001)Dogic, and Fraden]{DogicFraden2001}
	Dogic,~Z.; Fraden,~S. Development of model colloidal liquid crystals and the kinetics of the isotropic--smectic transition. \emph{Phil. Trans. Roy. Soc. A: Math. Phys. Engin. Sci.} \textbf{2001}, \emph{359}, 997--1015\relax
	\mciteBstWouldAddEndPuncttrue
	\mciteSetBstMidEndSepPunct{\mcitedefaultmidpunct}
	{\mcitedefaultendpunct}{\mcitedefaultseppunct}\relax
	\EndOfBibitem
	\bibitem[Bakker \latin{et~al.}(2016)Bakker, Dussi, Droste, Kennedy, Wiegant, Liu, Imhof, Dijkstra, and van Blaaderen]{Bakker2016}
	Bakker,~H.~E.; Dussi,~S.; Droste,~T.~H.,~Barbera L.and~Besseling; Kennedy,~C.~L.; Wiegant,~E.~I.; Liu,~B.; Imhof,~A.; Dijkstra,~M.; van Blaaderen,~A. Phase diagram of binary colloidal rod-sphere mixtures from a 3D real-space analysis of sedimentation–diffusion equilibria. \emph{Soft Matter} \textbf{2016}, \emph{12}, 9238--9245\relax
	\mciteBstWouldAddEndPuncttrue
	\mciteSetBstMidEndSepPunct{\mcitedefaultmidpunct}
	{\mcitedefaultendpunct}{\mcitedefaultseppunct}\relax
	\EndOfBibitem
	\bibitem[Peters \latin{et~al.}(2020)Peters, Vis, Wensink, and Tuinier]{PetersPRE2020}
	Peters,~V. F.~D.; Vis,~M.; Wensink,~H.~H.; Tuinier,~R. Algebraic equations of state for the liquid crystalline phase behavior of hard rods. \emph{Phys. Rev. E} \textbf{2020}, \emph{101}, 062707\relax
	\mciteBstWouldAddEndPuncttrue
	\mciteSetBstMidEndSepPunct{\mcitedefaultmidpunct}
	{\mcitedefaultendpunct}{\mcitedefaultseppunct}\relax
	\EndOfBibitem
	\bibitem[Lopes \latin{et~al.}(2021)Lopes, Romano, Grelet, Franco, and Giacometti]{Lopes2021}
	Lopes,~J.~T.; Romano,~F.; Grelet,~E.; Franco,~L. F.~M.; Giacometti,~A. Phase behavior of hard cylinders. \emph{J. Chem. Phys.} \textbf{2021}, \emph{154}, 104902\relax
	\mciteBstWouldAddEndPuncttrue
	\mciteSetBstMidEndSepPunct{\mcitedefaultmidpunct}
	{\mcitedefaultendpunct}{\mcitedefaultseppunct}\relax
	\EndOfBibitem
	\bibitem[Sawyer and Jaffe(1986)Sawyer, and Jaffe]{Sawyer1986}
	Sawyer,~L.; Jaffe,~M. The structure of thermotropic copolyesters. \emph{J. Mater. Sci.} \textbf{1986}, \emph{21}, 1897--1913\relax
	\mciteBstWouldAddEndPuncttrue
	\mciteSetBstMidEndSepPunct{\mcitedefaultmidpunct}
	{\mcitedefaultendpunct}{\mcitedefaultseppunct}\relax
	\EndOfBibitem
	\bibitem[Lu and Hsieh(2010)Lu, and Hsieh]{Lu2010}
	Lu,~P.; Hsieh,~Y.-L. Preparation and properties of cellulose nanocrystals: Rods, spheres, and network. \emph{Carbohydr. Pol.} \textbf{2010}, \emph{82}, 329--336\relax
	\mciteBstWouldAddEndPuncttrue
	\mciteSetBstMidEndSepPunct{\mcitedefaultmidpunct}
	{\mcitedefaultendpunct}{\mcitedefaultseppunct}\relax
	\EndOfBibitem
	\bibitem[Salas \latin{et~al.}(2014)Salas, Nypelo, Rodriguez-Abreu, Carrillo, and Rojas]{Salas2014}
	Salas,~C.; Nypelo,~T.; Rodriguez-Abreu,~C.; Carrillo,~C.; Rojas,~O.~J. Nanocellulose properties and applications in colloids and interfaces. \emph{Curr. Opin. Colloid Interface Sci.} \textbf{2014}, \emph{19}, 383--396\relax
	\mciteBstWouldAddEndPuncttrue
	\mciteSetBstMidEndSepPunct{\mcitedefaultmidpunct}
	{\mcitedefaultendpunct}{\mcitedefaultseppunct}\relax
	\EndOfBibitem
	\bibitem[Lagerwall \latin{et~al.}(2014)Lagerwall, Sch{\"u}tz, Salajkova, Noh, Hyun~Park, Scalia, and Bergstr{\"o}m]{Lagerwall2014}
	Lagerwall,~J.~P.; Sch{\"u}tz,~C.; Salajkova,~M.; Noh,~J.; Hyun~Park,~J.; Scalia,~G.; Bergstr{\"o}m,~L. Cellulose nanocrystal-based materials: from liquid crystal self-assembly and glass formation to multifunctional thin films. \emph{NPG Asia Materials} \textbf{2014}, \emph{6}, e80--e80\relax
	\mciteBstWouldAddEndPuncttrue
	\mciteSetBstMidEndSepPunct{\mcitedefaultmidpunct}
	{\mcitedefaultendpunct}{\mcitedefaultseppunct}\relax
	\EndOfBibitem
	\bibitem[Honorato-Rios \latin{et~al.}(2018)Honorato-Rios, Lehr, Sch{\"u}tz, Sanctuary, Osipov, Baller, and Lagerwall]{Honorato2018}
	Honorato-Rios,~C.; Lehr,~C.; Sch{\"u}tz,~C.; Sanctuary,~R.; Osipov,~M.~A.; Baller,~J.; Lagerwall,~J. P.~F. Fractionation of cellulose nanocrystals: enhancing liquid crystal ordering without promoting gelation. \emph{NPG Asia Materials} \textbf{2018}, \emph{10}, 455–465\relax
	\mciteBstWouldAddEndPuncttrue
	\mciteSetBstMidEndSepPunct{\mcitedefaultmidpunct}
	{\mcitedefaultendpunct}{\mcitedefaultseppunct}\relax
	\EndOfBibitem
	\bibitem[Thompson and Odijk(2004)Thompson, and Odijk]{ThompsonOdijk2004}
	Thompson,~J. M.~T.; Odijk,~T. Statics and dynamics of condensed DNA within phages and globules. \emph{Philos. Trans. Royal Soc. A} \textbf{2004}, \emph{362}, 1497--1517\relax
	\mciteBstWouldAddEndPuncttrue
	\mciteSetBstMidEndSepPunct{\mcitedefaultmidpunct}
	{\mcitedefaultendpunct}{\mcitedefaultseppunct}\relax
	\EndOfBibitem
	\bibitem[Roos \latin{et~al.}(2007)Roos, Ivanovska, Evilevitch, and Wuite]{Roos2007}
	Roos,~W.~H.; Ivanovska,~I.~L.; Evilevitch,~A.; Wuite,~G. J.~L. Viral capsids: Mechanical characteristics, genome packaging and delivery mechanisms. \emph{Cell. Mol. Life Sci.} \textbf{2007}, \emph{64}, 1484--1497\relax
	\mciteBstWouldAddEndPuncttrue
	\mciteSetBstMidEndSepPunct{\mcitedefaultmidpunct}
	{\mcitedefaultendpunct}{\mcitedefaultseppunct}\relax
	\EndOfBibitem
	\bibitem[Woldringh and Odijk(1999)Woldringh, and Odijk]{Woldringh1999}
	Woldringh,~C.~L.; Odijk,~T. In \emph{Organization of the prokaryotic genome}; Charlebois,~R.~L., Ed.; ASM Press: Amsterdam, 1999; Chapter 10, pp 171--187\relax
	\mciteBstWouldAddEndPuncttrue
	\mciteSetBstMidEndSepPunct{\mcitedefaultmidpunct}
	{\mcitedefaultendpunct}{\mcitedefaultseppunct}\relax
	\EndOfBibitem
	\bibitem[Khmelinskaia \latin{et~al.}(2021)Khmelinskaia, Franquelim, Yaadav, Petrov, and Schwille]{Khmelinskaia2021}
	Khmelinskaia,~A.; Franquelim,~H.~G.; Yaadav,~R.; Petrov,~E.~P.; Schwille,~P. Membrane-mediated self-organization of rod-like DNA origami on supported lipid bilayers. \emph{Adv. Mater. Interf.} \textbf{2021}, \emph{8}, 2101094\relax
	\mciteBstWouldAddEndPuncttrue
	\mciteSetBstMidEndSepPunct{\mcitedefaultmidpunct}
	{\mcitedefaultendpunct}{\mcitedefaultseppunct}\relax
	\EndOfBibitem
	\bibitem[Osborn and Rothfield(2007)Osborn, and Rothfield]{Osborn2007}
	Osborn,~M.~J.; Rothfield,~L. Cell shape determination in Escherichia coli. \emph{Current Opin. Microbiol.} \textbf{2007}, \emph{10}, 606--610\relax
	\mciteBstWouldAddEndPuncttrue
	\mciteSetBstMidEndSepPunct{\mcitedefaultmidpunct}
	{\mcitedefaultendpunct}{\mcitedefaultseppunct}\relax
	\EndOfBibitem
	\bibitem[Young(2007)]{Young2007}
	Young,~K.~D. {Bacterial morphology: Why have different shapes?} \emph{Current Opin. Microbiol.} \textbf{2007}, \emph{10}, 596--600\relax
	\mciteBstWouldAddEndPuncttrue
	\mciteSetBstMidEndSepPunct{\mcitedefaultmidpunct}
	{\mcitedefaultendpunct}{\mcitedefaultseppunct}\relax
	\EndOfBibitem
	\bibitem[Yang \latin{et~al.}(2016)Yang, Blair, and Salama]{Yang2016}
	Yang,~D.~C.; Blair,~K.~M.; Salama,~N.~R. Staying in shape: the impact of cell shape on bacterial survival in diverse environments. \emph{Microbiol. Mol. Biol. Rev.} \textbf{2016}, \emph{80}, 187--203\relax
	\mciteBstWouldAddEndPuncttrue
	\mciteSetBstMidEndSepPunct{\mcitedefaultmidpunct}
	{\mcitedefaultendpunct}{\mcitedefaultseppunct}\relax
	\EndOfBibitem
	\bibitem[Eichhorn \latin{et~al.}(2010)Eichhorn, Dufresne, Aranguren, Marcovich, Capadona, Rowan, Weder, Thielemans, Roman, Renneckar, Gindl, Veigel, Keckes, Yano, Abe, Nogi, Nakagaito, Mangalam, Simonsen, Benight, Bismarck, Berglund, and Peijs]{Eichhorn2010}
	Eichhorn,~S.~J.; Dufresne,~A.; Aranguren,~M.; Marcovich,~N.~E.; Capadona,~J.~R.; Rowan,~S.~J.; Weder,~C.; Thielemans,~W.; Roman,~M.; Renneckar,~S. \latin{et~al.}  Review: Current international research into cellulose nanofibres and nanocomposites. \emph{J. Mater. Sci.} \textbf{2010}, \emph{45}, 1--33\relax
	\mciteBstWouldAddEndPuncttrue
	\mciteSetBstMidEndSepPunct{\mcitedefaultmidpunct}
	{\mcitedefaultendpunct}{\mcitedefaultseppunct}\relax
	\EndOfBibitem
	\bibitem[Habibi \latin{et~al.}(2010)Habibi, Lucia, and Rojas]{Habibi2010}
	Habibi,~Y.; Lucia,~L.~A.; Rojas,~O.~J. Cellulose nanocrystals: Chemistry, self-assembly, and applications. \emph{Chem. Rev.} \textbf{2010}, \emph{110}, 3479--3500\relax
	\mciteBstWouldAddEndPuncttrue
	\mciteSetBstMidEndSepPunct{\mcitedefaultmidpunct}
	{\mcitedefaultendpunct}{\mcitedefaultseppunct}\relax
	\EndOfBibitem
	\bibitem[Palmer and Stupp(2008)Palmer, and Stupp]{Stupp2008}
	Palmer,~L.~C.; Stupp,~S.~I. Molecular self-assembly into one-dimensional nanostructures. \emph{Acc. Chem. Res.} \textbf{2008}, \emph{41}, 1674--1684\relax
	\mciteBstWouldAddEndPuncttrue
	\mciteSetBstMidEndSepPunct{\mcitedefaultmidpunct}
	{\mcitedefaultendpunct}{\mcitedefaultseppunct}\relax
	\EndOfBibitem
	\bibitem[Fraden \latin{et~al.}(1989)Fraden, Maret, Casper, and Meyer]{Fraden1989}
	Fraden,~S.; Maret,~G.; Casper,~D. L.~D.; Meyer,~R.~B. Isotropic-nematic phase transition and angular correlations in isotropic suspensions of tobacco mosaic virus. \emph{Phys. Rev. Lett.} \textbf{1989}, \emph{63}, 2068--2071\relax
	\mciteBstWouldAddEndPuncttrue
	\mciteSetBstMidEndSepPunct{\mcitedefaultmidpunct}
	{\mcitedefaultendpunct}{\mcitedefaultseppunct}\relax
	\EndOfBibitem
	\bibitem[Wen \latin{et~al.}(1989)Wen, Meyer, and Caspar]{Wen1989}
	Wen,~X.; Meyer,~R.~B.; Caspar,~D. L.~D. Observation of smectic-A ordering in a solution of rigid-rod-like particles. \emph{Phys. Rev. Lett.} \textbf{1989}, \emph{63}, 2760--2763\relax
	\mciteBstWouldAddEndPuncttrue
	\mciteSetBstMidEndSepPunct{\mcitedefaultmidpunct}
	{\mcitedefaultendpunct}{\mcitedefaultseppunct}\relax
	\EndOfBibitem
	\bibitem[Dogic and Fraden(1997)Dogic, and Fraden]{Dogic1997}
	Dogic,~Z.; Fraden,~S. Smectic phase in a colloidal suspension of semiflexible virus particles. \emph{Phys. Rev. Lett.} \textbf{1997}, \emph{78}, 2417--2420\relax
	\mciteBstWouldAddEndPuncttrue
	\mciteSetBstMidEndSepPunct{\mcitedefaultmidpunct}
	{\mcitedefaultendpunct}{\mcitedefaultseppunct}\relax
	\EndOfBibitem
	\bibitem[Grelet(2014)]{Grelet2014}
	Grelet,~E. Hard-rod behavior in dense mesophases of semiflexible and rigid charged viruses. \emph{Phys. Rev. X} \textbf{2014}, \emph{4}, 021053\relax
	\mciteBstWouldAddEndPuncttrue
	\mciteSetBstMidEndSepPunct{\mcitedefaultmidpunct}
	{\mcitedefaultendpunct}{\mcitedefaultseppunct}\relax
	\EndOfBibitem
	\bibitem[Janmey \latin{et~al.}(2014)Janmey, Slochower, Wang, Wen, and Cēbers]{Janmey2014}
	Janmey,~P.~A.; Slochower,~D.~R.; Wang,~Y.-H.; Wen,~Q.; Cēbers,~A. Polyelectrolyte properties of filamentous biopolymers and their consequences in biological fluids. \emph{Soft Matter} \textbf{2014}, \emph{10}, 1439--1449\relax
	\mciteBstWouldAddEndPuncttrue
	\mciteSetBstMidEndSepPunct{\mcitedefaultmidpunct}
	{\mcitedefaultendpunct}{\mcitedefaultseppunct}\relax
	\EndOfBibitem
	\bibitem[Tarafder \latin{et~al.}(2020)Tarafder, von K{\"u}gelgen, Mellul, Schulze, Aarts, and Bharat]{Tarafder2020}
	Tarafder,~A.~K.; von K{\"u}gelgen,~A.; Mellul,~A.~J.; Schulze,~U.; Aarts,~D. G. A.~L.; Bharat,~T. A.~M. Phage liquid crystalline droplets form occlusive sheaths that encapsulate and protect infectious rod-shaped bacteria. \emph{Proc. Natl. Acad. Sci. U.S.A.} \textbf{2020}, \emph{117}, 4724--4731\relax
	\mciteBstWouldAddEndPuncttrue
	\mciteSetBstMidEndSepPunct{\mcitedefaultmidpunct}
	{\mcitedefaultendpunct}{\mcitedefaultseppunct}\relax
	\EndOfBibitem
	\bibitem[Chapot \latin{et~al.}(2005)Chapot, Bocquet, and Trizac]{Chapot2005}
	Chapot,~D.; Bocquet,~L.; Trizac,~E. Electrostatic potential around charged finite rodlike macromolecules: Nonlinear Poisson–Boltzmann theory. \emph{J. Colloid Interface Sci.} \textbf{2005}, \emph{285}, 609--618\relax
	\mciteBstWouldAddEndPuncttrue
	\mciteSetBstMidEndSepPunct{\mcitedefaultmidpunct}
	{\mcitedefaultendpunct}{\mcitedefaultseppunct}\relax
	\EndOfBibitem
	\bibitem[Verwey and Overbeek(1948)Verwey, and Overbeek]{VerweyOverbeek1948}
	Verwey,~E. J.~W.; Overbeek,~J.~T. \emph{Theory of the stability of lyophobic colloids}; Elsevier: Amsterdam, 1948; pp 1--205\relax
	\mciteBstWouldAddEndPuncttrue
	\mciteSetBstMidEndSepPunct{\mcitedefaultmidpunct}
	{\mcitedefaultendpunct}{\mcitedefaultseppunct}\relax
	\EndOfBibitem
	\bibitem[Adams and Fraden(1998)Adams, and Fraden]{Adams1998}
	Adams,~M.; Fraden,~S. {Phase behavior of mixtures of rods (tobacco mosaic virus) and spheres (polyethylene oxide, bovine serum albumin)}. \emph{Biophys. J.} \textbf{1998}, \emph{74}, 669--677\relax
	\mciteBstWouldAddEndPuncttrue
	\mciteSetBstMidEndSepPunct{\mcitedefaultmidpunct}
	{\mcitedefaultendpunct}{\mcitedefaultseppunct}\relax
	\EndOfBibitem
	\bibitem[Adams \latin{et~al.}(1998)Adams, Dogic, Keller, and Fraden]{Adams1998a}
	Adams,~M.; Dogic,~Z.; Keller,~S.~L.; Fraden,~S. Entropically driven microphase transitions in mixtures of colloidal rods and spheres. \emph{Nature} \textbf{1998}, \emph{393}, 349--352\relax
	\mciteBstWouldAddEndPuncttrue
	\mciteSetBstMidEndSepPunct{\mcitedefaultmidpunct}
	{\mcitedefaultendpunct}{\mcitedefaultseppunct}\relax
	\EndOfBibitem
	\bibitem[Dogic \latin{et~al.}(2000)Dogic, Frenkel, and Fraden]{Dogic2000}
	Dogic,~Z.; Frenkel,~D.; Fraden,~S. Enhanced stability of layered phases in parallel hard spherocylinders due to addition of hard spheres. \emph{Phys. Rev. E.} \textbf{2000}, \emph{62}, 3925--3933\relax
	\mciteBstWouldAddEndPuncttrue
	\mciteSetBstMidEndSepPunct{\mcitedefaultmidpunct}
	{\mcitedefaultendpunct}{\mcitedefaultseppunct}\relax
	\EndOfBibitem
	\bibitem[Dogic and Fraden(2001)Dogic, and Fraden]{Dogic2001}
	Dogic,~Z.; Fraden,~S. Development of model colloidal liquid crystals and the kinetics of the isotropic--smectic transition. \emph{Phil. Trans. R. Soc. Lond. A} \textbf{2001}, \emph{359}, 997--1015\relax
	\mciteBstWouldAddEndPuncttrue
	\mciteSetBstMidEndSepPunct{\mcitedefaultmidpunct}
	{\mcitedefaultendpunct}{\mcitedefaultseppunct}\relax
	\EndOfBibitem
	\bibitem[Dogic and Fraden(2006)Dogic, and Fraden]{Dogic2006b}
	Dogic,~Z.; Fraden,~S. \emph{Soft Matter: Complex Colloidal Suspensions}; John Wiley \& Sons, Ltd, 2006; Vol.~2; Chapter 1, pp 1--86\relax
	\mciteBstWouldAddEndPuncttrue
	\mciteSetBstMidEndSepPunct{\mcitedefaultmidpunct}
	{\mcitedefaultendpunct}{\mcitedefaultseppunct}\relax
	\EndOfBibitem
	\bibitem[Honorato-Rios \latin{et~al.}(2016)Honorato-Rios, Kuhnhold, Bruckner, Dannert, Schilling, and Lagerwall]{Honorato2016}
	Honorato-Rios,~C.; Kuhnhold,~A.; Bruckner,~J.~R.; Dannert,~R.; Schilling,~T.; Lagerwall,~J. P.~F. Equilibrium liquid crystal phase diagrams and detection of kinetic arrest in cellulose nanocrystal suspensions. \emph{Front. Mater.} \textbf{2016}, 21\relax
	\mciteBstWouldAddEndPuncttrue
	\mciteSetBstMidEndSepPunct{\mcitedefaultmidpunct}
	{\mcitedefaultendpunct}{\mcitedefaultseppunct}\relax
	\EndOfBibitem
	\bibitem[Stroobants \latin{et~al.}(1986)Stroobants, Lekkerkerker, and Odijk]{SLO1986}
	Stroobants,~A.; Lekkerkerker,~H. N.~W.; Odijk,~T. Effect of electrostatic interaction on the liquid crystal phase transition in solutions of rodlike polyelectrolytes. \emph{Macromolecules} \textbf{1986}, \emph{19}, 2232--2238\relax
	\mciteBstWouldAddEndPuncttrue
	\mciteSetBstMidEndSepPunct{\mcitedefaultmidpunct}
	{\mcitedefaultendpunct}{\mcitedefaultseppunct}\relax
	\EndOfBibitem
	\bibitem[Vologodskii and Cozzarelli(1995)Vologodskii, and Cozzarelli]{Vologodskii1995}
	Vologodskii,~A.; Cozzarelli,~N. Modeling of long-range electrostatic interactions in DNA. \emph{Biopolymers} \textbf{1995}, \emph{35}, 289--296\relax
	\mciteBstWouldAddEndPuncttrue
	\mciteSetBstMidEndSepPunct{\mcitedefaultmidpunct}
	{\mcitedefaultendpunct}{\mcitedefaultseppunct}\relax
	\EndOfBibitem
	\bibitem[Widom(1963)]{Widom1963}
	Widom,~B. Some topics in the theory of fluids. \emph{J. Chem. Phys.} \textbf{1963}, \emph{39}, 2808--2812\relax
	\mciteBstWouldAddEndPuncttrue
	\mciteSetBstMidEndSepPunct{\mcitedefaultmidpunct}
	{\mcitedefaultendpunct}{\mcitedefaultseppunct}\relax
	\EndOfBibitem
	\bibitem[V{\"o}rtler and Smith(2000)V{\"o}rtler, and Smith]{Vortler2000}
	V{\"o}rtler,~H.~L.; Smith,~W.~R. Computer simulation studies of a square-well fluid in a slit pore. Spreading pressure and vapor–liquid phase equilibria using the virtual-parameter-variation method. \emph{J. Chem. Phys.} \textbf{2000}, \emph{112}, 5168--5174\relax
	\mciteBstWouldAddEndPuncttrue
	\mciteSetBstMidEndSepPunct{\mcitedefaultmidpunct}
	{\mcitedefaultendpunct}{\mcitedefaultseppunct}\relax
	\EndOfBibitem
	\bibitem[Cotter(1974)]{Cotter1974}
	Cotter,~M.~A. Hard-rod fluid: Scaled particle theory revisited. \emph{Phys. Rev. A} \textbf{1974}, \emph{10}, 625--636\relax
	\mciteBstWouldAddEndPuncttrue
	\mciteSetBstMidEndSepPunct{\mcitedefaultmidpunct}
	{\mcitedefaultendpunct}{\mcitedefaultseppunct}\relax
	\EndOfBibitem
	\bibitem[McGrother \latin{et~al.}(1996)McGrother, Williamson, and Jackson]{McGrother1996}
	McGrother,~S.~C.; Williamson,~D.~C.; Jackson,~G. A re‐examination of the phase diagram of hard spherocylinders. \emph{J. Chem. Phys.} \textbf{1996}, \emph{104}, 6755--6771\relax
	\mciteBstWouldAddEndPuncttrue
	\mciteSetBstMidEndSepPunct{\mcitedefaultmidpunct}
	{\mcitedefaultendpunct}{\mcitedefaultseppunct}\relax
	\EndOfBibitem
	\bibitem[Opdam \latin{et~al.}(2021)Opdam, Guu, Schelling, Aarts, Tuinier, and Lettinga]{Opdam2021b}
	Opdam,~J.; Guu,~D.; Schelling,~M. P.~M.; Aarts,~D. G. A.~L.; Tuinier,~R.; Lettinga,~M.~P. Phase stability of colloidal mixtures of spheres and rods. \emph{J. Chem. Phys.} \textbf{2021}, \emph{154}, 204906\relax
	\mciteBstWouldAddEndPuncttrue
	\mciteSetBstMidEndSepPunct{\mcitedefaultmidpunct}
	{\mcitedefaultendpunct}{\mcitedefaultseppunct}\relax
	\EndOfBibitem
	\bibitem[Philip and Wooding(1970)Philip, and Wooding]{PhilipWooding1970}
	Philip,~J.~R.; Wooding,~R.~A. Solution of the Poisson–Boltzmann equation about a cylindrical particle. \emph{J. Chem. Phys.} \textbf{1970}, \emph{52}, 953--959\relax
	\mciteBstWouldAddEndPuncttrue
	\mciteSetBstMidEndSepPunct{\mcitedefaultmidpunct}
	{\mcitedefaultendpunct}{\mcitedefaultseppunct}\relax
	\EndOfBibitem
	\bibitem[Hill(1955)]{Hill1955}
	Hill,~T.~L. Approximate calculation of the electrostatic free energy of nucleic acids and other cylindrical macromolecules. \emph{Arch. Biochem. Biophys.} \textbf{1955}, \emph{57}, 229--239\relax
	\mciteBstWouldAddEndPuncttrue
	\mciteSetBstMidEndSepPunct{\mcitedefaultmidpunct}
	{\mcitedefaultendpunct}{\mcitedefaultseppunct}\relax
	\EndOfBibitem
	\bibitem[Vroege and Lekkerkerker(1992)Vroege, and Lekkerkerker]{Vroege1992}
	Vroege,~G.~J.; Lekkerkerker,~H. N.~W. Phase transitions in lyotropic colloidal and polymer liquid crystals. \emph{Rep. Progr. Phys.} \textbf{1992}, \emph{55}, 1241\relax
	\mciteBstWouldAddEndPuncttrue
	\mciteSetBstMidEndSepPunct{\mcitedefaultmidpunct}
	{\mcitedefaultendpunct}{\mcitedefaultseppunct}\relax
	\EndOfBibitem
	\bibitem[Opdam \latin{et~al.}(2022)Opdam, Gandhi, Kuhnhold, Schilling, and Tuinier]{Opdam2022}
	Opdam,~J.; Gandhi,~P.; Kuhnhold,~A.; Schilling,~T.; Tuinier,~R. Excluded volume interactions and phase stability in mixtures of hard spheres and hard rods. \emph{Phys. Chem. Chem. Phys.} \textbf{2022}, \emph{24}, 11820--11827\relax
	\mciteBstWouldAddEndPuncttrue
	\mciteSetBstMidEndSepPunct{\mcitedefaultmidpunct}
	{\mcitedefaultendpunct}{\mcitedefaultseppunct}\relax
	\EndOfBibitem
	\bibitem[Odijk(1986)]{Odijk1986}
	Odijk,~T. Theory of Lyotropic Polymer Liquid Crystals. \emph{Macromolecules} \textbf{1986}, \emph{19}, 2313--2329\relax
	\mciteBstWouldAddEndPuncttrue
	\mciteSetBstMidEndSepPunct{\mcitedefaultmidpunct}
	{\mcitedefaultendpunct}{\mcitedefaultseppunct}\relax
	\EndOfBibitem
	\bibitem[Drwenski \latin{et~al.}(2016)Drwenski, Dussi, Hermes, Dijkstra, and van Roij]{Drwenski2016}
	Drwenski,~T.; Dussi,~S.; Hermes,~M.; Dijkstra,~M.; van Roij,~R. {Phase diagrams of charged colloidal rods: Can a uniaxial charge distribution break chiral symmetry?} \emph{J. Chem. Phys.} \textbf{2016}, \emph{144}, 094901\relax
	\mciteBstWouldAddEndPuncttrue
	\mciteSetBstMidEndSepPunct{\mcitedefaultmidpunct}
	{\mcitedefaultendpunct}{\mcitedefaultseppunct}\relax
	\EndOfBibitem
	\bibitem[Han and Herzfeld(1996)Han, and Herzfeld]{Han1996}
	Han,~J.; Herzfeld,~J. An avoidance model for short-range order induced by soft repulsions in systems of rigid rods. \emph{MRS Online Proceedings Library (OPL)} \textbf{1996}, \emph{463}, 135--140\relax
	\mciteBstWouldAddEndPuncttrue
	\mciteSetBstMidEndSepPunct{\mcitedefaultmidpunct}
	{\mcitedefaultendpunct}{\mcitedefaultseppunct}\relax
	\EndOfBibitem
	\bibitem[Kramer and Herzfeld(1999)Kramer, and Herzfeld]{Kramer1999}
	Kramer,~E.~M.; Herzfeld,~J. Avoidance model for soft particles. I. Charged spheres and rods beyond the dilute limit. \emph{J. Chem. Phys.} \textbf{1999}, \emph{110}, 8825--8834\relax
	\mciteBstWouldAddEndPuncttrue
	\mciteSetBstMidEndSepPunct{\mcitedefaultmidpunct}
	{\mcitedefaultendpunct}{\mcitedefaultseppunct}\relax
	\EndOfBibitem
\end{mcitethebibliography}

\providecommand{\latin}[1]{#1}
\makeatletter
\providecommand{\doi}
{\begingroup\let\do\@makeother\dospecials
	\catcode`\{=1 \catcode`\}=2 \doi@aux}
\providecommand{\doi@aux}[1]{\endgroup\texttt{#1}}
\makeatother
\providecommand*\mcitethebibliography{\thebibliography}
\csname @ifundefined\endcsname{endmcitethebibliography}  {\let\endmcitethebibliography\endthebibliography}{}

%\clearpage

\clearpage 
\renewcommand{\thefigure}{S\arabic{figure}}
\renewcommand{\thepage}{S\arabic{page}} 
\setcounter{figure}{0}
\setcounter{page}{1}
\begin{suppinfo}
Here we show the relative difference between the approximation and the exact expression for $A'$, and some computer simulation results for other rod aspect ratios as those presented in the main text.
%\end{suppinfo}
%\begin{suppinfo}

\begin{figure}[bp!]
\centering
\includegraphics[trim=0 0 0 0,clip,width=0.7\columnwidth]{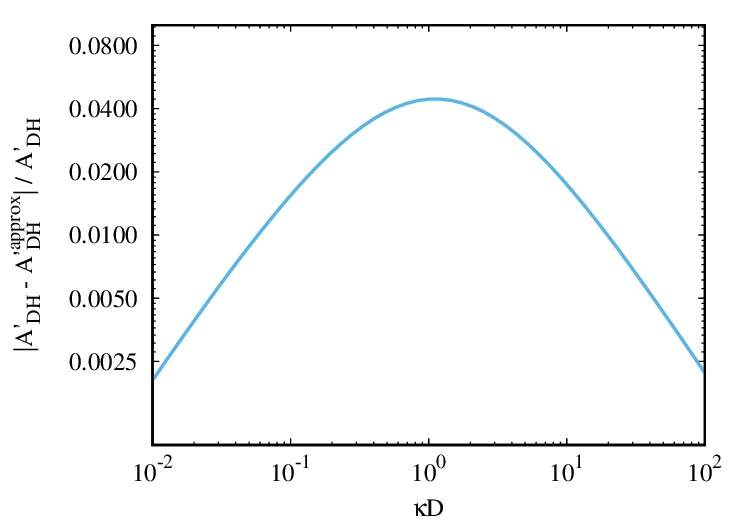}
\caption{Relative difference between Eqs.~%(11) and (13)
(\ref{AprimeeqSLO}) and (\ref{AprimeeqSLOapprox})
.}
\label{SIe}
\end{figure}

\begin{figure*}[bh!]
\centering
\begin{picture}(470,310)%(0.99\columnwidth,0.5\textheight)
\put(10,155){\includegraphics[trim=0 0 0 0,clip,width=0.5\columnwidth]{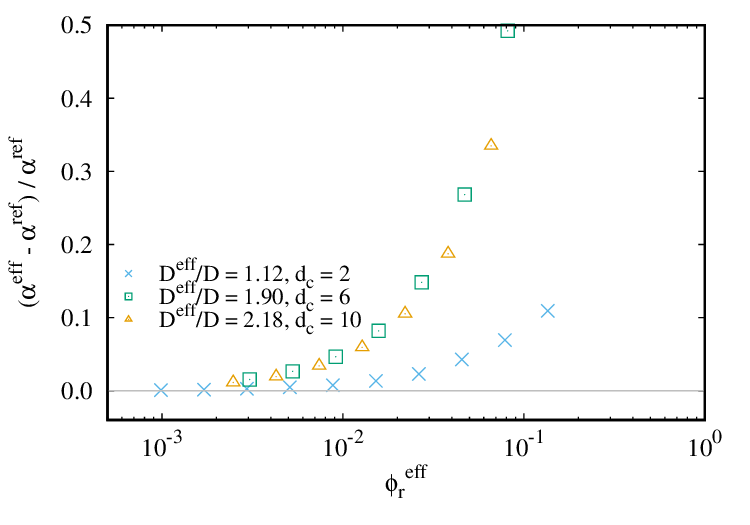}}
\put(50,242){\includegraphics[trim=0 0 0 0,clip,width=0.2\columnwidth]{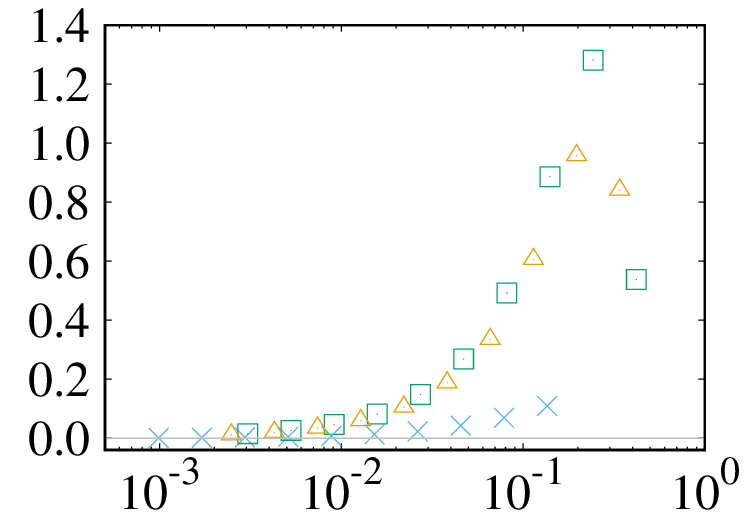}}
\put(240,155){\includegraphics[trim=0 0 0 0,clip,width=0.5\columnwidth]{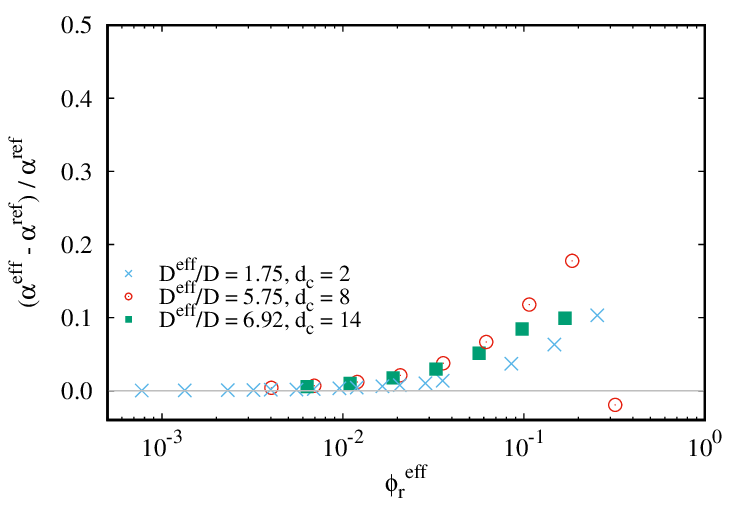}}
\put(10,0){\includegraphics[trim=0 0 0 0,clip,width=0.5\columnwidth]{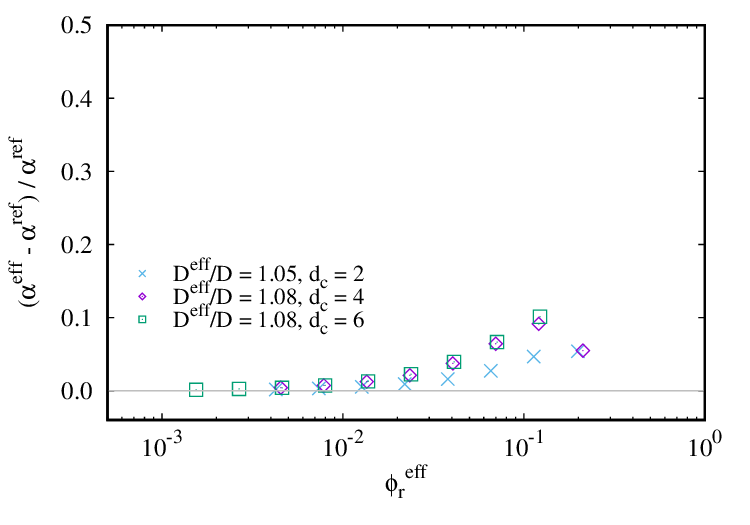}}
\put(240,0){\includegraphics[trim=0 0 0 0,clip,width=0.5\columnwidth]{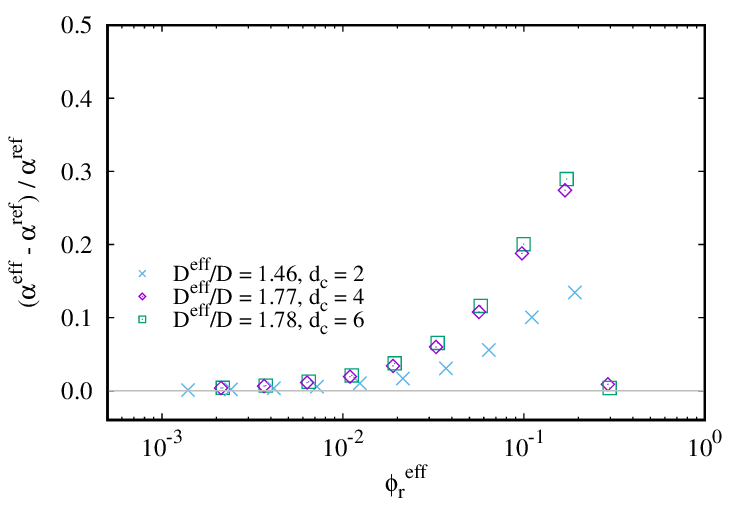}}
\put(10,298){\small{\textbf{(a)}}}
\put(240,298){\small{\textbf{(b)}}}
\put(10,144){\small{\textbf{(c)}}}
\put(240,144){\small{\textbf{(d)}}}
\end{picture}
\caption{Relative difference between the effective free volume fraction and the reference from the immersion free energy vs.\ effective volume fraction for different cutoff lengths as indicated in the legend and $(A,\kappa D)=$ $(1.0,0.5)$(a), $(16.0,0.5)$(b), $(1.0,2.0)$(c), $(16.0,2.0)$(d). The hard rod aspect ratio is $L/D+1=6$ and the effective diameter ratio $D^\mathrm{eff}/D$ is indicated in the legend.}
\label{SIa}
\end{figure*}

\begin{figure*}[tbp]
\centering
\begin{picture}(470,310)%(0.99\columnwidth,0.5\textheight)
\put(10,155){\includegraphics[trim=0 0 0 0,clip,width=0.5\columnwidth]{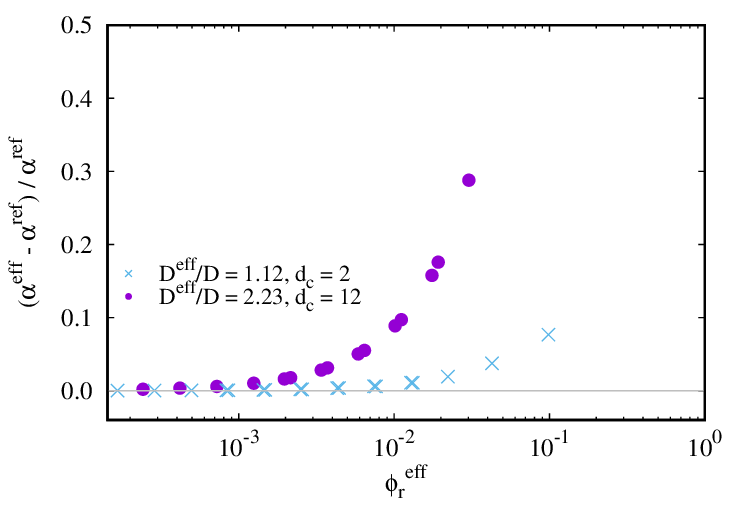}}
\put(50,242){\includegraphics[trim=0 0 0 0,clip,width=0.2\columnwidth]{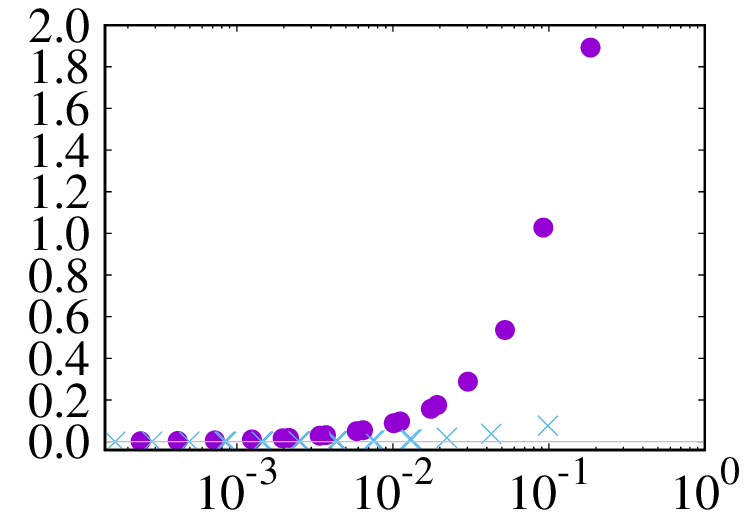}}
\put(240,155){\includegraphics[trim=0 0 0 0,clip,width=0.5\columnwidth]{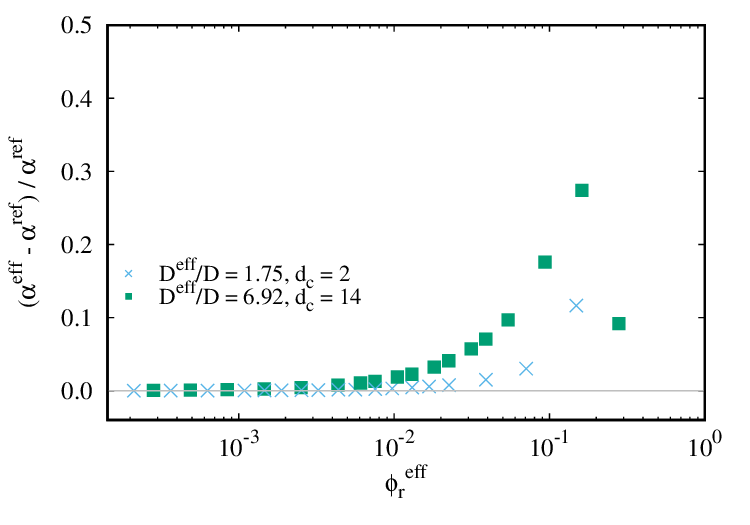}}
\put(10,0){\includegraphics[trim=0 0 0 0,clip,width=0.5\columnwidth]{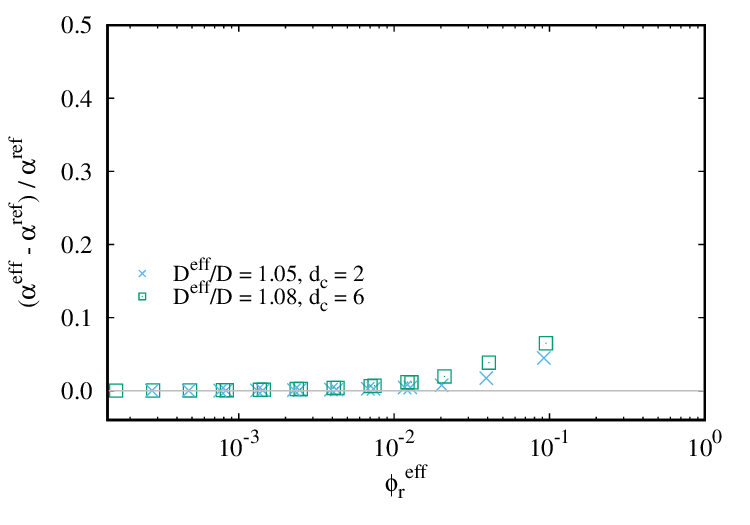}}
\put(240,0){\includegraphics[trim=0 0 0 0,clip,width=0.5\columnwidth]{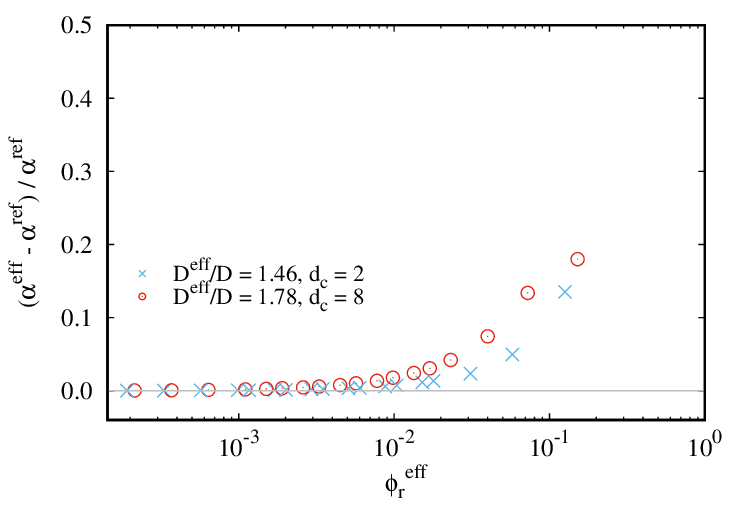}}
\put(10,298){\small{\textbf{(a)}}}
\put(240,298){\small{\textbf{(b)}}}
\put(10,144){\small{\textbf{(c)}}}
\put(240,144){\small{\textbf{(d)}}}
\end{picture}
\caption{Relative difference between the effective free volume fraction and the reference from the immersion free energy vs.\ effective volume fraction for different cutoff lengths as indicated in the legend and $(A,\kappa D)=$ $(1.0,0.5)$(a), $(16.0,0.5)$(b), $(1.0,2.0)$(c), $(16.0,2.0)$(d). The hard rod aspect ratio is $L/D+1=21$ and the effective diameter ratio $D^\mathrm{eff}/D$ is indicated in the legend.}
\label{SIc}
\end{figure*}

\begin{figure*}[tb]
\centering
\begin{picture}(470,155)%(0.99\columnwidth,0.5\textheight)
\put(10,0){\includegraphics[trim=0 0 0 0,clip,width=0.5\columnwidth]{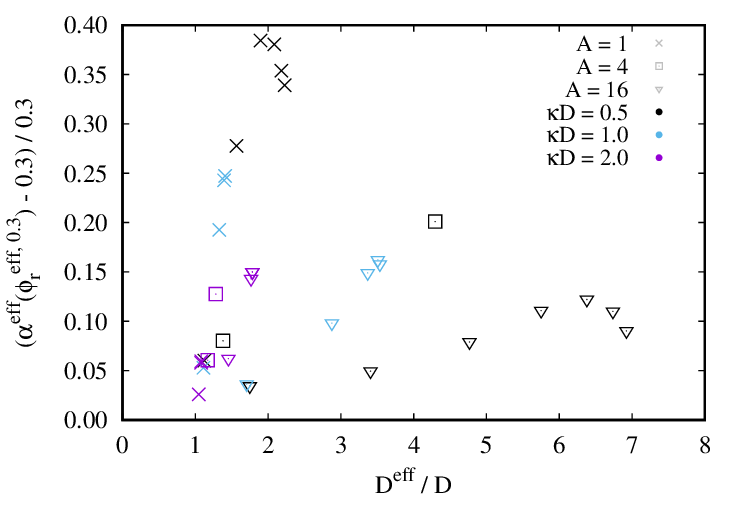}}
\put(240,0){\includegraphics[trim=0 0 0 0,clip,width=0.5\columnwidth]{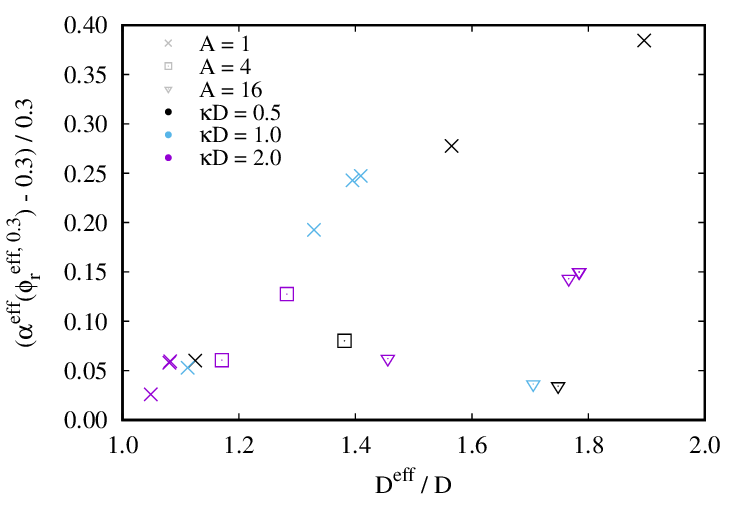}}
\put(10,144){\small{\textbf{(a)}}}
\put(240,144){\small{\textbf{(b)}}}
\end{picture}
\caption{Relative difference between the effective free volume fraction and the reference from the immersion free energy at the effective volume fraction for which the reference equals 0.3 vs.\ effective diameter for different $A$ and $\kappa D$. Identical symbols refer to the same $(A,\kappa D)$ but different cutoff lengths (resulting in different effective diameter ratios. The hard rod aspect ratio is $L/D+1=6$. (a) full range of effective diameters. (b) small effective diameters.}
\label{SIb}
\end{figure*}

\begin{figure*}[tbp]
\centering
\begin{picture}(470,155)%(0.99\columnwidth,0.5\textheight)
\put(10,0){\includegraphics[trim=0 0 0 0,clip,width=0.5\columnwidth]{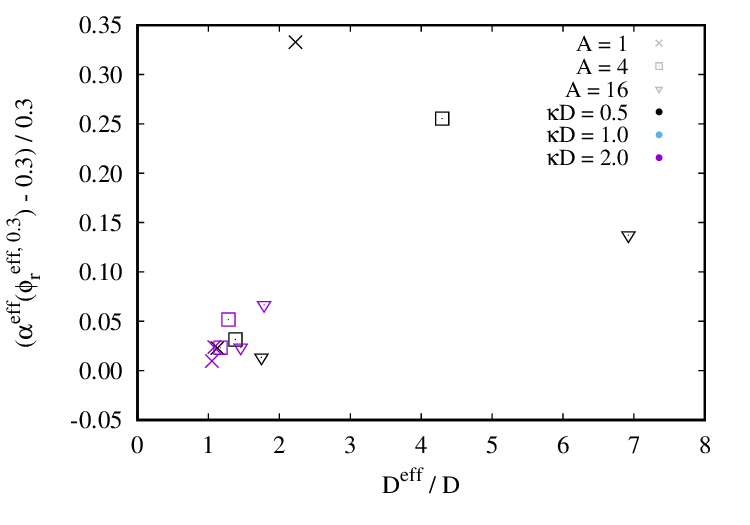}}
\put(240,0){\includegraphics[trim=0 0 0 0,clip,width=0.5\columnwidth]{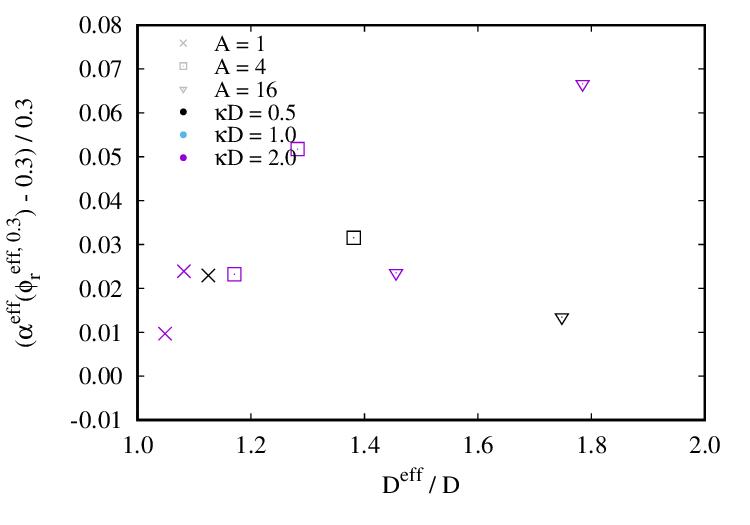}}
\put(10,144){\small{\textbf{(a)}}}
\put(240,144){\small{\textbf{(b)}}}
\end{picture}
\caption{Relative difference between the effective free volume fraction and the reference from the immersion free energy at the effective volume fraction for which the reference equals 0.3 vs.\ effective diameter for different $A$ and $\kappa D$. Identical symbols refer to the same $(A,\kappa D)$ but different cutoff lengths (resulting in different effective diameter ratios. The hard rod aspect ratio is $L/D+1=21$. (a) full range of effective diameters. (b) small effective diameters.}
\label{SId}
\end{figure*}

\end{suppinfo}

\end{document}